\documentclass[11pt]{article}%
\usepackage{amsmath}
\usepackage{graphicx}%
\usepackage{amsfonts}%
\usepackage{amssymb}
\newcommand{\eqnb}{\begin{equation}}
\newcommand{\eqne}{\end{equation}}

\newtheorem{The}{Theorem}

\newtheorem{Lem}{Lemma}

\newtheorem{Rem}{Remark}

\begin{document}

\title{Groups of Repairmen and Repair-based Load Balancing in Supermarket Models with
Repairable Servers}
\author{Na Li\\Department of Industrial Engineering and Management\\Shanghai Jiaotong University, Shanghai 200240, China\\Quan-Lin Li\\School of Economics and Management Sciences\\Yanshan University, Qinhuangdao 066004, China \\Zhe George Zhang \\Department of Decision Sciences, Western Washington University, \\Beedie School of Business, Simon Fraser University}
\maketitle

\begin{abstract}
Supermarket models are a class of interesting parallel queueing
networks with dynamic randomized load balancing and real-time
resource management. When the parallel servers are subject to
breakdowns and repairs, analysis of such a supermarket model becomes
more difficult and challenging. In this paper, we apply the
mean-field theory to studying four interrelated supermarket models
with repairable servers, and numerically indicate impact of the
different repairman groups on performance of the systems. First, we
set up the systems of mean-field equations for the supermarket
models with repairable servers. Then we prove the asymptotic
independence of the supermarket models through the operator
semigroup and the mean-field limit. Furthermore, we show that the
fixed points of the supermarket models satisfy the systems of
nonlinear equations. Finally, we use the fixed points to give
numerical computation for performer analysis, and provide valuable
observations on model improvement. Therefore, this paper provides a
new and effective method in the study of complex supermarket models.

\vskip                                          0.5cm


\end{abstract}

\section{Introduction}

In the last two decades considerable research attention has been paid to the
study of supermarket models. Supermarket models are a class of interesting
parallel queueing networks with dynamic and real-time adaptive control, for
example, size-based routine selection, and information-based resource
scheduling. Such a supermarket model can be applied to, such as, computer
networks, manufacturing systems, transportation networks and healthcare
systems. Since a simple supermarket model was discussed by Mitzenmacher
\cite{Mit:1996}, Vvedenskaya et al. \cite{Vve:1996} and Turner \cite{Tur:1996,
Tur:1998}, more studies have been done by, for instance, Vvedenskaya and Suhov
\cite{Vve:1997}, Graham \cite{Gra:2000, Gra:2004}, Luczak and McDiarmid
\cite{Luc:2006, Luc:2007}, Bramson et al. \cite{Bra:2010, Bra:2012, Bra:2013},
Li \cite{Li:2014}, Li et al. \cite{Li:2014a, Li:2015} and Li and Lui
\cite{Li:2016}, Gast et al. \cite{Gast:2011}, Gast and Gaujal \cite{Gast:2012}
and Mukhopadhyay et al. \cite{Muk:2015}. For the fast Jackson networks (or
supermarket networks), readers may refer to Martin and Suhov \cite{Mar:1999},
Martin \cite{Mar:2001} and Suhov and Vvedenskaya \cite{Suh:2002}.

In many stochastic networks, servers subject to breakdowns and repairs always
encounter in practical areas, such as, computer systems, communication
networks, manufacturing systems, and transportation networks. Because system
performance deteriorates quickly due to servers' breakdowns and limited repair
capacity, analyzing such a stochastic systems with repairable servers is not
only important from theoretical perspective but also necessary from practical
engineering. On this research line, important examples include Mitrany and
Avi-Ttzhak \cite{Mit:1968}, Neuts and Lucantoni \cite{Neu:1979}, Kulkarni and
Choi \cite{Kul:1990}, Li et al. \cite{LiW:1997}, Aissani and Artalejo
\cite{Ais:1998}, N\'{u}\~{n}ez-Queija \cite{Nun:2000}, Li et al.
\cite{Li:2006}, Economou and Kantaa \cite{Eco:2008}, Fiems et al.
\cite{Fie:2008}, Kamoun \cite{Kam:2008}, and Krishnamoorthy et al.
\cite{Kri:2012} for a survey.

It is interesting but difficult to discuss stochastic systems of parallel
queues with unreliable servers, e.g., see an excellent survey by Adan et al.
\cite{Ada:2001}. Up to now, the available results on parallel-queue systems
with repairable servers are still very few. Andradottir et al. \cite{And:2007}
applied a Markov decision process to compensating for failures with flexible
servers. Martonosi \cite{Mart:2011} studied a dynamic server allocation at
parallel queues with unreliable servers. Saghafian et al. \cite{Sag:2011}
analyzed the dynamic control of unreliable flexible servers in a ``W''
network. Ravid et al. \cite{Rav:2013} considered the repair systems with
exchangeable items and the longest queue mechanism. Stimulated by practical
need of many distributed parallel systems, the study of supermarket models and
work stealing models is highly paid attention on computer systems and
communication networks. This motivates us in this paper to apply the
mean-field theory to analyzing supermarket models with servers subject to
breakdowns and repairs, which are a class of important complex reliability
networks, and specifically, the different groups of repairmen make analysis of
such a reliability network more difficult and challenging.

It is necessary to provide a simple survey for the mean-field theory. The
mean-field equations and the asymptotic independence (or propagation of chaos)
play an important role in the study of interacting particle systems, e.g., see
Liggett \cite{Lig:2012} and Kipnis and Landim \cite{Kip:2013}. For the
mean-field theory of complex stochastic systems, readers may refer to, for
example, interacting Markov processes by Spitzer \cite{Spi:1970}, Dawson
\cite{Daw:1983}, Sznitman \cite{Szn:1989} and Chen \cite{Chen:2004} and Li
\cite{Li:2016a}; queueing networks by Baccelli et al. \cite{Bac:1992},
Borovkov \cite{Bor:1998} and Mitzenmacher et al. \cite{Mit:2001}; work
stealing models by Gast and Gaujal \cite{Gast:2010} and Li and Yang
\cite{Li:2015a}; communication networks by Duffield \cite{Duf:1992}, Benaim
and Le Boudec \cite{Ben:2008}, Duffy \cite{Duf:2010} and Bordenave et al.
\cite{Bor:2010}.

The main contributions of this paper are threefold. The first one is to
describe and analyze a class of important complex reliability networks:
Supermarket models with repairable servers, which play a key role in
performance evaluation of computer systems and of communication networks.
Notice that a supermarket model contains multiple repairable servers, thus the
different groups of repairmen make analysis of the supermarket model more
complicated. In the situation, this paper considers four interrelated
supermarket models with repairable servers through observing two different
arrival dispatched schemes and two different groups of repairmen. The second
contribution is to apply the mean-field theory to studying the four
interrelated supermarket models with repairable servers. This paper
demonstrates such a mean-field analysis through the following three steps: (a)
Providing a probability computation for setting up the systems of mean-field
equations, (b) calculating the fixed points through the systems of nonlinear
equations, and (c) giving performance analysis of the supermarket models with
repairable servers and developing numerical computation for useful observation
on model improvement. The third contribution is to provide a better example in
order to demonstrate how to develop numerical solution in the study of complex
supermarket models. Since the nonlinear structure of the mean-field equations
makes a supermarket model almost impossible to find an analytic solution to
the system of mean-field equations, it is a key to sufficiently develop
numerical computation in performance evaluation of supermarket models. Based
on this, numerical examples are used to provide valuable observations on how
to improve performance of supermarket models either from system parameter
optimization or from various resource deployment (e.g., arrival dispatched
schemes, allocated service ability, and groups of repairmen).

Finally, note that this paper discusses a special class of supermarket models
with unreliable servers, while their failed states and the groups of repairmen
have influence on the arrival joining schemes. To analyze such a supermarket
model, the most relevant references to this paper are Li et al.
\cite{Li:2014a, Li:2015} and Li and Lui \cite{Li:2016} from two points of
view: (1) The environment invariant factors were proposed to setting up
systems of mean-field equations for complex supermarket models. As studied in
Li et al. \cite{Li:2015}, this paper also analyzes a double dynamic routine
selection scheme both for the arrival dispatched schemes and for the groups of
repairmen. It is worthwhile to note that such a multiple dynamic routine
selection scheme is a new and interesting topic in the study of supermarket
models and of work stealing models.

The remainder of this paper is organized as follows. In Section 2, we first
describe four interrelated supermarket models with repairable servers where
customer arrivals make use of system information and repair ability is grouped
in some different structures. Then we use the fraction vector to describe an
infinite-dimensional Markov process for each supermarket model with repairable
servers. In Sections 3, we provide two types of probability representations
both for the arrival dispatched schemes by means of system information
and\ for the repair ability grouped in different ways. In Sections 4, for each
of the four interrelated supermarket models with repairable servers, we set up
an infinite-dimensional system of mean-field equations. In Section 5, we
discuss the fixed points for the systems of mean-field equations, and show
that the fixed points can be determined by the systems of nonlinear equations.
In Section 6, we first provide useful performance measures of the supermarket
models with repairable servers. Then we use some numerical examples to make
valuable observations on model improvement by means of performance numerical
comparison. Section 7 concludes with a summary. The proofs of some key results
are provided in Appendix A.

\section{Supermarket Models with Repairable Servers}

In this section, we first describe four interrelated supermarket models with
repairable servers, where the arrival dispatched schemes make use of system
information and the repair ability is grouped in different ways. Then we use
the fraction vector (or empirical measure) to describe an infinite-dimensional
Markov process for each supermarket model with repairable servers.

\subsection{Model description}

\textbf{The arrival processes}

Customers arrive at the system as a Poisson process with arrival rate
$N\lambda$ for $\lambda>0$. Upon arrival, an arriving customer chooses
$d_{1}(\geq1)$ servers from the $N$ servers independently and randomly. Then
the customer will select one server (or queue) to join. Such a server
selection is based on two different information observations as follows:

\textbf{(A.1)} \underline{\textit{Observing only the shortest queue}%
}\textit{.} The arriving customer joins the shortest queue among the $d_{1}$
queues. If there is a tie, the customer makes the choice equally likely among
the shortest queues of the same length.

\textbf{(A.2)} \underline{\textit{Observing both the shortest queue
and the status (working or repairing) of the }}
\newline\underline{$d_{1}$ \textit{selected servers}}. The arriving
customer joins the shortest queue with the working server as higher
priority than the server in repair among the $d_{1}$ selected
servers.

\textbf{The service processes }

The service times at each server are i.i.d. and are exponential distributed
with service rate $\mu>0$.

\textbf{The repair processes }

Each server has an exponential life time with failure rate $\alpha>0$. When
the server fails, it enters a failure state and undergoes the repair process
immediately. The service of a customer interrupted by a server's failure is
resumed as soon as the server is repaired. We assume that the repaired server
is as good as new and the service time is cumulative. To deploy the repair
resource effectively, we consider three types of repair schemes as follows:

\textbf{(R.1) }\underline{\textit{Each server has one repairman}}\textit{.}
There are $N$ repairmen corresponding to the $N$ servers, and thus each server
has a repairman of itself. The repair times are i.i.d exponential random
variables with repair rate $\beta$.

\textbf{(R.2) }\underline{\textit{A super large
repairman}}\textit{.}\textbf{ }There is only one fast repairman
whose repair time is exponentially distributed with repair rate
$N\beta$ and $\beta>0$. This super repairman chooses $d_{2}(\geq1)$
servers from the $N$ servers randomly. If all the $d_{2}$ servers
are working, then the repairman is idle; if at least one of the
selected $d_{2}$ servers is failed, then the repairman repairs the
failed server with the longest queue. If there is a tie, the
repairman select one randomly.

\textbf{(R.3) }\underline{\textit{A large repairman and }$J$\textit{ small
repairmen for }$0\leq J\leq N-1$}\textit{. }There are a large repairman and
$J$ small repairmen, where the repair time of the large repairman is
exponentially distributed with the repair rate $\left(  N-J\right)  \beta$,
and the repair time of each small repairman is exponentially distributed with
repair rate $\beta$.

Each of the $J$ small repairmen can repair one failed server at a time, if
any; whilst the large repairman chooses $d_{2}$ servers from the $N$ servers
independently and randomly. If all the selected $d_{2}$ servers are working,
then the large repairman is idle; if at least one of the selected $d_{2}$
servers is failed but not repaired by small repairmen yet, then the large
repairman repairs the failed server with the longest queue. If there is a tie,
the repairman selects the failed server with the longest queue.

We assume that all the random variables defined above are independent of each
other. Figure 1 shows a supermarket model with repairable servers and a large repairman.

\begin{figure}[ptbh]
\centering            \includegraphics[width=12cm]{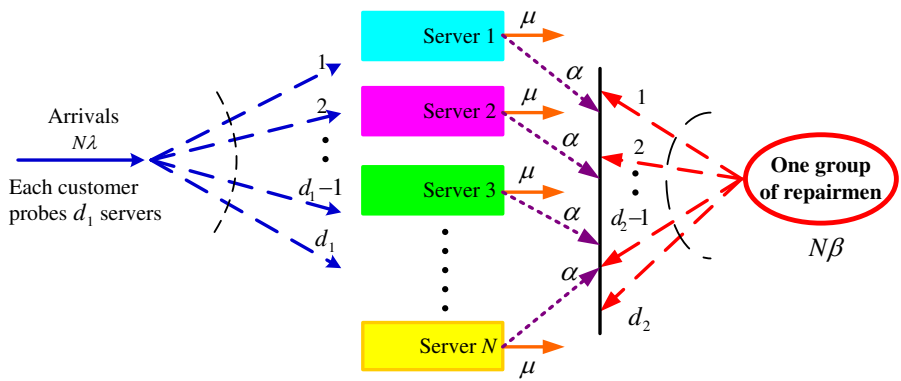}  \caption{A
physical illustration of a supermarket model with repairable servers}%
\label{figure: fig-1}%
\end{figure}

Now, we construct four interrelated supermarket models with repairable
servers, which are constructed by different combinations of (A.$i$)\textbf{
}and (R.$i$) for $i=1,2$ as follows:

\textbf{Model I} ((A.$1$)\textbf{ }and (R.$1$)): In this model, an arriving
customer only needs to observe the queue lengths of the $d_{1}$ selected
server and joins the shortest queue. There are $N$ repairmen corresponding to
the $N$ servers, hence each server has one repairman of itself.

\textbf{Model II} ((A.$1$)\textbf{ }and (R.$2$)): In this model, the queue
selection rule is the same as Model I. However, there is only one super
repairman who chooses $d_{2}$ servers from the $N$ servers independently and
uniformly at random. If all the selected $d_{2}$ servers are working, then the
repairman is idle; otherwise, the repairman repairs the failed server with the
longest queue length.

\textbf{Model III} ((A.$2$)\textbf{ }and (R.$1$)): In this model, an arriving
customer observe not only the queue lengths of the $d_{1}$ selected servers,
but also the states (working or repairing) of the $d_{1}$ selected servers.
The customer then joins the shortest queue with working servers having higher
priority than failed servers. There are $N$ repairmen corresponding to the $N$
servers, hence each server has one repairman of itself.

\textbf{Model IV} ((A.$2$)\textbf{ }and (R.$2$)): In this model, the
customer's queue selection rule is the same as Model III. However, there is
only a super repairman, which chooses $d_{2}$ servers from the $N$ servers
independently and uniformly at random. If all the selected $d_{2}$ servers are
working, then the repairman is idle; otherwise, the repairman repairs the
failed server with the longest queue.

\begin{Rem}
Actually, (R.$3$) is a more general scheme of repair resource allocation, and
its analysis can be completed through by modifying the mean-field equations in
(R.$1$) and (R.$2$). Here, we do not consider (R.$3$), and (R.$3$) will be
investigated in another paper.
\end{Rem}

Next, we shall provide a complete mathematical analysis for the four
interrelated supermarket models, and present some numerical examples to show
how the system information ((A.$i$)\textbf{ } and repair resource allocation
(R.$i$) for $i=1,2$) affect performance of the supermarket models with
repairable servers. Some insightful observations are made for designing and
controlling the arrival, service and repair processes to improve the
supermarket models.

\subsection{An infinite-dimensional Markov process}

Now, we use the empirical measure to provide an infinite-dimensional Markov
process for studying each of the four interrelated supermarket models with
repairable servers.

For $k\geq0$, we denote by $n_{k}^{\left(  W\right)  }\left(  t\right)  $ the
numbers of working (or idle) servers\ with at least $k\geq0$ customers at time
$t\geq0$, and $n_{l}^{\left(  R\right)  }\left(  t\right)  $ the numbers of
failed servers with at least $l\geq1$ customers\ at time $t\geq0$. Clearly,
$n_{0}^{\left(  W\right)  }\left(  t\right)  +n_{1}^{\left(  R\right)
}\left(  t\right)  =N$ and $0\leq n_{k}^{\left(  W\right)  }\left(  t\right)
,n_{l}^{\left(  R\right)  }\left(  t\right)  \leq N$ for $k\geq0$ and $l\geq1$.

We write that for $k\geq0$,
\[
U_{W,k}^{\left(  N\right)  }\left(  t\right)  =\frac{n_{k}^{\left(  W\right)
}\left(  t\right)  }{N}%
\]
and for $l\geq1$%
\[
V_{R,l}^{\left(  N\right)  }\left(  t\right)  =\frac{n_{l}^{\left(  R\right)
}\left(  t\right)  }{N},
\]
which are the fractions of working (or idle) servers with at least $k$
customers and of failed servers with at least $l$ customers at time $t\geq0$,
respectively. Let%
\[
\mathbf{U}_{W}^{\left(  N\right)  }\left(  t\right)  =\left(  U_{W,0}^{\left(
N\right)  }\left(  t\right)  ,U_{W,1}^{\left(  N\right)  }\left(  t\right)
,U_{W,2}^{\left(  N\right)  }\left(  t\right)  ,\ldots\right)  ,
\]%
\[
\mathbf{V}_{R}^{\left(  N\right)  }\left(  t\right)  =\left(  V_{R,1}^{\left(
N\right)  }\left(  t\right)  ,V_{R,2}^{\left(  N\right)  }\left(  t\right)
,V_{R,3}^{\left(  N\right)  }\left(  t\right)  ,\ldots\right)  ,
\]
and%
\[
\mathbf{U}^{\left(  N\right)  }\left(  t\right)  =\left(  \mathbf{U}%
_{W}^{\left(  N\right)  }\left(  t\right)  ,\mathbf{V}_{R}^{\left(  N\right)
}\left(  t\right)  \right)  .
\]
Clearly, the state of the supermarket model of $N$ identical repairable
servers is described as a stochastic process $\left\{  \mathbf{U}^{\left(
N\right)  }\left(  t\right)  :t\geq0\right\}  $. Since the arrival process is
Poisson, and the distributions of the service, life and repair times are all
exponential, $\left\{  \mathbf{U}^{\left(  N\right)  }\left(  t\right)
:t\geq0\right\}  $ is an infinite-dimensional Markov process whose state space
is given by%
\begin{align*}
\mathbf{E}_{N}=  &  \left\{  \left(  \mathbf{u}^{\left(  N\right)
},\mathbf{v}^{\left(  N\right)  }\right)  :1\geq u_{0}^{\left(  N\right)
}\geq u_{1}^{\left(  N\right)  }\geq u_{2}^{\left(  N\right)  }\geq
u_{3}^{\left(  N\right)  }\geq\cdots\geq0,\right. \\
&  1\geq v_{1}^{\left(  N\right)  }\geq v_{2}^{\left(  N\right)  }\geq
v_{3}^{\left(  N\right)  }\geq v_{4}^{\left(  N\right)  }\geq\cdots\geq0,\\
&  \left.  Nu_{k}^{\left(  N\right)  }\text{ and }Nv_{l}^{\left(  N\right)
}\text{ are nonnegative integers for }k\geq0\text{ and }l\geq1\right\}  .
\end{align*}

For a fixed pair array $\left(  t,N\right)  $ with $t\geq0$ and
$N=1,2,3,\ldots$, it is easy to see from the stochastic order that
$U_{W,k}^{\left(  N\right)  }\left(  t\right)  \geq U_{W,k+1}^{\left(
N\right)  }\left(  t\right)  $\ for $k\geq0$ and $V_{R,l}^{\left(  N\right)
}\left(  t\right)  \geq V_{R,l+1}^{\left(  N\right)  }\left(  t\right)  $\ for
$l\geq1$. This gives
\begin{equation}
1\geq U_{W,0}^{\left(  N\right)  }\left(  t\right)  \geq U_{W,1}^{\left(
N\right)  }\left(  t\right)  \geq U_{W,2}^{\left(  N\right)  }\left(
t\right)  \geq U_{W,3}^{\left(  N\right)  }\left(  t\right)  \geq\cdots\geq0
\label{Inequ-01}%
\end{equation}
and%
\begin{equation}
1\geq V_{R,1}^{\left(  N\right)  }\left(  t\right)  \geq V_{R,2}^{\left(
N\right)  }\left(  t\right)  \geq V_{R,3}^{\left(  N\right)  }\left(
t\right)  \geq V_{R,4}^{\left(  N\right)  }\left(  t\right)  \geq\cdots\geq0.
\label{Inequ-02}%
\end{equation}

To study the infinite-dimensional Markov process $\left\{  \mathbf{U}^{\left(
N\right)  }\left(  t\right)  :t\geq0\right\}  $, we write the expected
fractions as follows%
\[
u_{W,k}^{\left(  N\right)  }\left(  t\right)  =E\left[  U_{W,k}^{\left(
N\right)  }\left(  t\right)  \right]
\]
and%
\[
u_{R,l}^{\left(  N\right)  }\left(  t\right)  =E\left[  V_{R,l}^{\left(
N\right)  }\left(  t\right)  \right]  .
\]
It is easy to see from (\ref{Inequ-01}) and (\ref{Inequ-02}) that%
\begin{equation}
1\geq u_{W,0}^{(N)}\left(  t\right)  \geq u_{W,1}^{(N)}\left(  t\right)  \geq
u_{W,2}^{(N)}\left(  t\right)  \geq u_{W,3}^{(N)}\left(  t\right)  \geq
\cdots\geq0 \label{Inequ-1}%
\end{equation}
and%
\begin{equation}
1\geq u_{R,1}^{\left(  N\right)  }\left(  t\right)  \geq u_{R,2}^{\left(
N\right)  }\left(  t\right)  \geq u_{R,3}^{\left(  N\right)  }\left(
t\right)  \geq u_{R,4}^{\left(  N\right)  }\left(  t\right)  \geq\cdots\geq0.
\label{Inequ-2}%
\end{equation}

Let%
\[
\mathbf{u}_{W}^{\left(  N\right)  }\left(  t\right)  =\left(  u_{W,0}^{\left(
N\right)  }\left(  t\right)  ,u_{W,1}^{\left(  N\right)  }\left(  t\right)
,u_{W,2}^{\left(  N\right)  }\left(  t\right)  ,u_{W,3}^{\left(  N\right)
}\left(  t\right)  ,\ldots\right)  ,
\]%
\[
\mathbf{V}_{R}^{\left(  N\right)  }\left(  t\right)  =\left(  u_{R,1}^{\left(
N\right)  }\left(  t\right)  ,u_{R,2}^{\left(  N\right)  }\left(  t\right)
,u_{R,3}^{\left(  N\right)  }\left(  t\right)  ,u_{R,4}^{\left(  N\right)
}\left(  t\right)  ,\ldots\right)
\]
and%
\[
\mathbf{u}^{\left(  N\right)  }\left(  t\right)  =\left(  \mathbf{u}%
_{W}^{\left(  N\right)  }\left(  t\right)  ,\mathbf{V}_{R}^{\left(  N\right)
}\left(  t\right)  \right)  .
\]

\section{Two Types of Probability Representations}

In this section, we provide two types of probability representations for
customer arrivals by means of system information and\ for repair ability
grouped in different ways. For notational simplicity, the two types of
probability representations are denoted as the four pair control schemes:
((A.$i$),\textbf{ }(R.$i$)) for $i=1,2$. The probability representations are
useful for establishing the systems of mean-field equations later.

For the supermarket models of $N$ identical repairable servers, to set up the
probability representations, we only need to determine the expected change in
the number of servers with at least $k$ customers over a small time period
$\left[  0,\text{d}t\right)  $.

\subsection{The arrival processes}

This subsection provides the probability representations for the arrival
processes, in which the two different cases of (A.$1$) and (A.$2$) are
discussed. Note that the analysis of (A.$1$) is similar to that of Li et al.
\cite{Li:2014a}. To make our paper self-contained, we still present some
computational details for (A.$1$) and (R.$1$). For (A.$2$) and (R.$2$), we
only provide the main results.

\textbf{(A.}$1$\textbf{): Observing the Queue Length Only}

To give the probability representations, we need to compute the rate that any
arriving customer selects $d_{1}$ servers from the $N$ servers independently
and uniformly at random, and joins the selected server with the shortest
queue. Note that the arriving customer does not have the server status
information (working or repair). Thus our computation for such a rate contains
two steps as follows:

\underline{Step I: Entering one working server}

In this step, the rate that any arriving customer joins a working server with
the shortest queue length $k-1$ is given by%
\begin{equation}
N\lambda\left[  u_{W,k-1}^{(N)}\left(  t\right)  -u_{W,k}^{(N)}\left(
t\right)  \right]  W_{W,k}^{\left(  N\right)  }\left(  u_{W,k-1}%
,u_{W,k};u_{R,k-1},u_{R,k};t\right)  \text{d}t, \label{Equ4}%
\end{equation}
where%
\begin{align}
&  W_{W,1}^{\left(  N\right)  }\left(  u_{W,0},u_{W,1};u_{R,1};t\right)
=\sum_{m=1}^{d_{1}}C_{d_{1}}^{m}\left[  u_{W,0}^{(N)}\left(  t\right)
-u_{W,1}^{(N)}\left(  t\right)  \right]  ^{m-1}\left[  u_{W,1}^{(N)}\left(
t\right)  \right]  ^{d_{1}-m}\nonumber\\
&  +\sum_{m=1}^{d_{1}-1}C_{d_{1}}^{m}\left[  u_{W,0}^{(N)}\left(  t\right)
-u_{W,1}^{(N)}\left(  t\right)  \right]  ^{m-1}\sum_{j=1}^{d_{1}-m}C_{d_{1}%
-m}^{j}\left[  u_{W,1}^{(N)}\left(  t\right)  \right]  ^{d_{1}-m-j}\left[
u_{R,1}^{(N)}\left(  t\right)  \right]  ^{j}, \label{Equ-4-1}%
\end{align}
and for $k\geq2$
\begin{align}
&  W_{W,k}^{\left(  N\right)  }\left(  u_{W,k-1},u_{W,k};u_{R,k-1}%
,u_{R,k};t\right)  =\sum_{m=1}^{d_{1}}C_{d_{1}}^{m}\left[  u_{W,k-1}%
^{(N)}\left(  t\right)  -u_{W,k}^{(N)}\left(  t\right)  \right]  ^{m-1}\left[
u_{W,k}^{(N)}\left(  t\right)  \right]  ^{d_{1}-m}\nonumber\\
&  +\sum_{m=1}^{d_{1}-1}C_{d_{1}}^{m}\left[  u_{W,k-1}^{(N)}\left(  t\right)
-u_{W,k}^{(N)}\left(  t\right)  \right]  ^{m-1}\sum_{j=1}^{d_{1}-m}C_{d_{1}%
-m}^{j}\left[  u_{W,k}^{(N)}\left(  t\right)  \right]  ^{d_{1}-m-j}\left[
u_{R,k}^{(N)}\left(  t\right)  \right]  ^{j}\nonumber\\
&  +\sum_{m=2}^{d_{1}}C_{d_{1}}^{m}\sum_{m_{1}=1}^{m-1}\frac{m_{1}}{m}%
C_{m}^{m_{1}}\left[  u_{W,k-1}^{(N)}\left(  t\right)  -u_{W,k}^{(N)}\left(
t\right)  \right]  ^{m_{1}-1}\nonumber\\
&  \times\left[  u_{R,k-1}^{(N)}\left(  t\right)  -u_{R,k}^{(N)}\left(
t\right)  \right]  ^{m-m_{1}}\sum_{r=0}^{d_{1}-m}C_{d_{1}-m}^{r}\left[
u_{W,k}^{(N)}\left(  t\right)  \right]  ^{r}\left[  u_{R,k}^{(N)}\left(
t\right)  \right]  ^{d_{1}-m-r}. \label{Equ-4-2}%
\end{align}

\begin{figure}[ptb]
\centering                   \includegraphics[width=10cm]{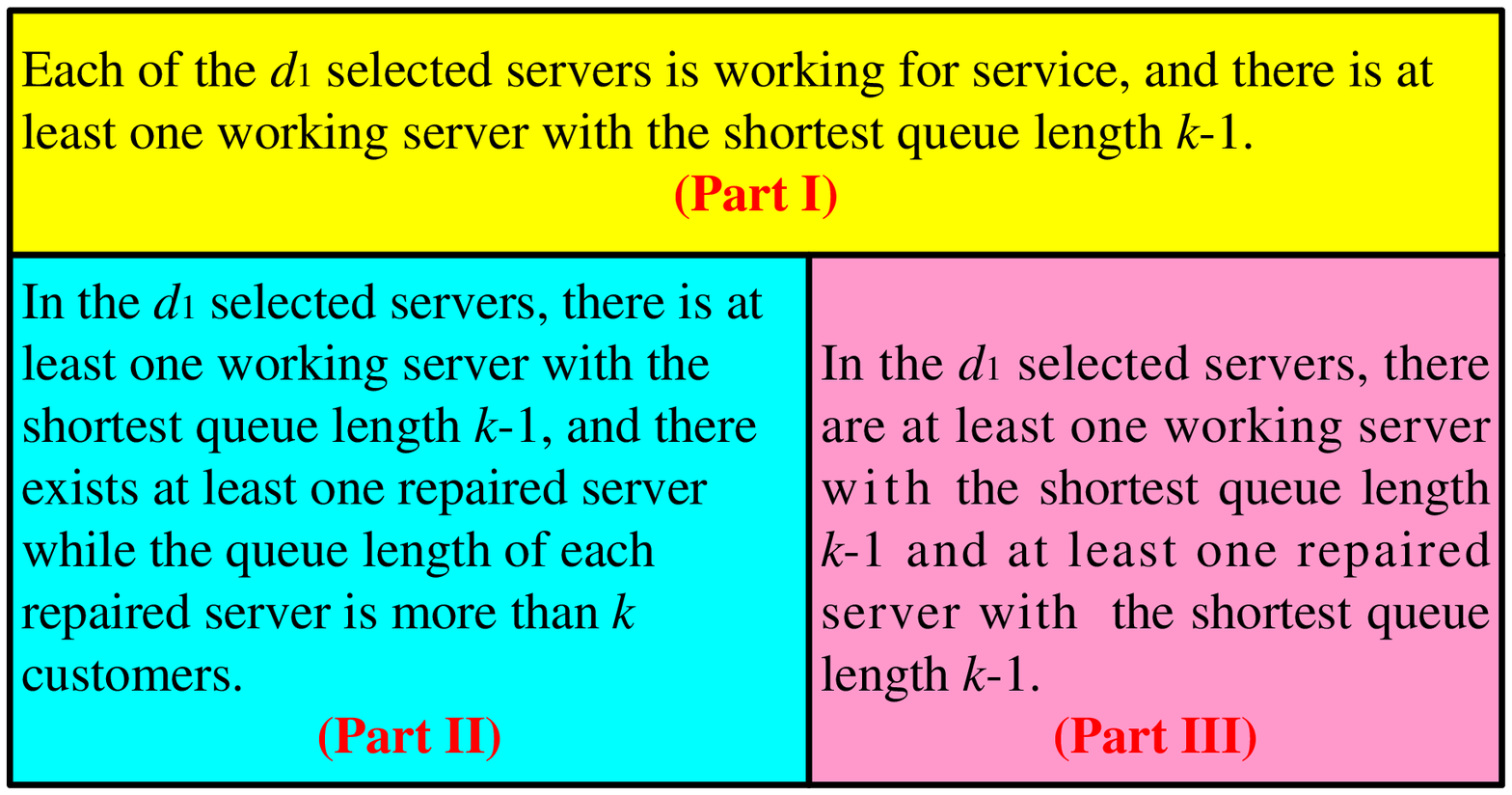}
\newline \caption{Set decomposition of possible events when joining a working
server}%
\label{figure:fig-2}%
\end{figure}

To derive the probabilities $W_{W,1}^{\left(  N\right)  }\left(
u_{W,0},u_{W,1};u_{R,1};t\right)  $ and $W_{W,k}^{\left(  N\right)  }\left(
u_{W,k-1},u_{W,k};u_{R,k-1},u_{R,k};t\right)  $ for $k\geq2$, Figure 2 shows
the set decomposition of all possible events, and the probabilities are
derived from the following three parts, that is,%
\[
W_{W,k}^{\left(  N\right)  }\left(  u_{W,k-1},u_{W,k};u_{R,k-1},u_{R,k}%
;t\right)  =\text{Part I }+\text{Part II }+\text{Part III}.
\]

\textit{Part I: None of the }$d_{1}$\textit{ selected servers is in repair}.
All $d_{1}$ selected servers are working for serving customers. In this case,
the probability that any arriving customer joins a working server with the
shortest queue length $k-1$ and the queue lengths of the other selected
$d_{1}-1$ working servers are not shorter than $k-1$ is given by%
\begin{align}
&  \sum_{m=1}^{d_{1}}C_{d_{1}}^{m}\left[  u_{W,k-1}^{(N)}\left(  t\right)
-u_{W,k}^{(N)}\left(  t\right)  \right]  ^{m}\left[  u_{W,k}^{(N)}\left(
t\right)  \right]  ^{d_{1}-m}\nonumber\\
&  =\left[  u_{W,k-1}^{(N)}\left(  t\right)  -u_{W,k}^{(N)}\left(  t\right)
\right]  \sum_{m=1}^{d_{1}}C_{d_{1}}^{m}\left[  u_{W,k-1}^{(N)}\left(
t\right)  -u_{W,k}^{(N)}\left(  t\right)  \right]  ^{m-1}\left[  u_{W,k}%
^{(N)}\left(  t\right)  \right]  ^{d_{1}-m}, \label{Equ-4-3}%
\end{align}
where $C_{d_{1}}^{m}=d_{1}!/\left[  m!\left(  d_{1}-m\right)  !\right]  $ is a
binomial coefficient, $\left[  u_{W,k-1}^{(N)}\left(  t\right)  -u_{W,k}%
^{(N)}\left(  t\right)  \right]  ^{m}$ is the probability that any arriving
customer who can only choose one queue makes $m$ independent selections during
the $m$ selected working servers with the queue length $k-1$ at time $t$, and
$\left[  u_{W,k}^{(N)}\left(  t\right)  \right]  ^{d-m}$ is the probability
that any arriving customer who can only choose one queue makes $d_{1}-m$
independent selections during the $d_{1}-m$ selected working servers whose
queue lengths are not shorter than $k$ at time $t$.

\textit{Part I\negthinspace I: For the }$d_{1}$\textit{ selected servers,
there is at least one working server with the shortest queue length }%
$k-1$\textit{, and there exist at least one server in repair while the queue
length of each server in repair is more than }$k$\textit{ customers}. In this
case, the probability that any arriving customer joins a working server with
the shortest queue length $k-1$; and for the other $d_{1}-1$ selected servers,
the queue lengths of the selected working servers are not shorter than $k-1$,
and there exist at least one server in repair while the queue length of each
server in repair is more than $k$ customers, is given by
\begin{align}
&  \sum_{m=1}^{d_{1}-1}C_{d_{1}}^{m}\left[  u_{W,k-1}^{(N)}\left(  t\right)
-u_{W,k}^{(N)}\left(  t\right)  \right]  ^{m}\sum_{j=1}^{d_{1}-m}C_{d_{1}%
-m}^{j}\left[  u_{W,k}^{(N)}\left(  t\right)  \right]  ^{d_{1}-m-j}\left[
u_{R,k}^{(N)}\left(  t\right)  \right]  ^{j}\nonumber\\
&  =\left[  u_{W,k-1}^{(N)}\left(  t\right)  -u_{W,k}^{(N)}\left(  t\right)
\right]  \sum_{m=1}^{d_{1}-1}C_{d_{1}}^{m}\left[  u_{W,k-1}^{(N)}\left(
t\right)  -u_{W,k}^{(N)}\left(  t\right)  \right]  ^{m-1}\nonumber\\
&  \times\sum_{j=1}^{d_{1}-m}C_{d_{1}-m}^{j}\left[  u_{W,k}^{(N)}\left(
t\right)  \right]  ^{d_{1}-m-j}\left[  u_{R,k}^{(N)}\left(  t\right)  \right]
^{j}. \label{Equ-4-4}%
\end{align}

\textit{Part I\negthinspace I\negthinspace I: For the }$d_{1}$\textit{
selected servers, there is at least one working server with the shortest queue
length }$k-1$\textit{ and there is at least one server in repair with the
shortest queue length }$k-1$. In this case, if there are the $m$ selected
servers with the shortest queue length $k-1$ where there are $m_{1}\ $working
servers and $m-m_{1}$ servers in repair, then the probability that any
arriving customer joins a working server is equal to $m_{1}/m$. Therefore, the
probability that any arriving customer joins a working server with the
shortest queue length $k-1$, the queue lengths of the other $d_{1}-1$ selected
servers are not shorter than $k-1$, and there are at least one working server
with $k-1$ customers and at least one server in repair with $k-1$ customers is
given by%
\begin{align}
&  \sum_{m=2}^{d_{1}}C_{d_{1}}^{m}\sum_{m_{1}=1}^{m-1}\frac{m_{1}}{m}%
C_{m}^{m_{1}}\left[  u_{W,k-1}^{(N)}\left(  t\right)  -u_{W,k}^{(N)}\left(
t\right)  \right]  ^{m_{1}}\left[  u_{R,k-1}^{(N)}\left(  t\right)
-u_{R,k}^{(N)}\left(  t\right)  \right]  ^{m-m_{1}}\nonumber\\
&  \times\sum_{r=0}^{d_{1}-m}C_{d_{1}-m}^{r}\left[  u_{W,k}^{(N)}\left(
t\right)  \right]  ^{r}\left[  u_{R,k}^{(N)}\left(  t\right)  \right]
^{d_{1}-m-r}\nonumber\\
&  =\left[  u_{W,k-1}^{(N)}\left(  t\right)  -u_{W,k}^{(N)}\left(  t\right)
\right]  \sum_{m=2}^{d_{1}}C_{d_{1}}^{m}\sum_{m_{1}=1}^{m-1}\frac{m_{1}}%
{m}C_{m}^{m_{1}}\left[  u_{W,k-1}^{(N)}\left(  t\right)  -u_{W,k}^{(N)}\left(
t\right)  \right]  ^{m_{1}-1}\nonumber\\
&  \times\left[  u_{R,k-1}^{(N)}\left(  t\right)  -u_{R,k}^{(N)}\left(
t\right)  \right]  ^{m-m_{1}}\sum_{r=0}^{d_{1}-m}C_{d_{1}-m}^{r}\left[
u_{W,k}^{(N)}\left(  t\right)  \right]  ^{r}\left[  u_{R,k}^{(N)}\left(
t\right)  \right]  ^{d_{1}-m-r}. \label{Equ-4-5}%
\end{align}

\underline{Step two: Entering one server in repair}

This step can be dealt with similarly to that in Step one. The rate that any
arriving customer joins one server in repair with the shortest queue length
$k-1$ and the queue lengths of the other selected $d_{1}-1$ servers are not
shorter than $k-1$ is given by%
\begin{equation}
N\lambda\left[  u_{R,k-1}^{(N)}\left(  t\right)  -u_{R.k}^{(N)}\left(
t\right)  \right]  W_{R,k}^{\left(  N\right)  }\left(  u_{W,k-1}%
,u_{W,k};u_{R,k-1},u_{R,k};t\right)  \text{d}t, \label{Equ6}%
\end{equation}
where%
\begin{align}
&  W_{R,k}^{\left(  N\right)  }\left(  u_{W,k-1},u_{W,k};u_{R,k-1}%
,u_{R,k};t\right)  =\sum_{m=1}^{d_{1}}C_{d_{1}}^{m}\left[  u_{R,k-1}%
^{(N)}\left(  t\right)  -u_{R.k}^{(N)}\left(  t\right)  \right]  ^{m-1}\left[
u_{R,k}^{(N)}\left(  t\right)  \right]  ^{d_{1}-m}\nonumber\\
&  +\sum_{m=1}^{d_{1}-1}C_{d_{1}}^{m}\left[  u_{R,k-1}^{(N)}\left(  t\right)
-u_{R.k}^{(N)}\left(  t\right)  \right]  ^{m-1}\sum_{j=1}^{d_{1}-m}C_{d_{1}%
-m}^{j}\left[  u_{R,k}^{(N)}\left(  t\right)  \right]  ^{d_{1}-m-j}\left[
u_{W,k}^{(N)}\left(  t\right)  \right]  ^{j}\nonumber\\
&  +\sum_{m=2}^{d_{1}}C_{d_{1}}^{m}\sum_{m_{1}=1}^{m-1}\frac{m_{1}}{m}%
C_{m}^{m_{1}}\left[  u_{R,k-1}^{(N)}\left(  t\right)  -u_{R.k}^{(N)}\left(
t\right)  \right]  ^{m_{1}-1}\nonumber\\
&  \times\left[  u_{W,k-1}^{(N)}\left(  t\right)  -u_{W,k}^{(N)}\left(
t\right)  \right]  ^{m-m_{1}}\sum_{r=0}^{d_{1}-m}C_{d_{1}-m}^{r}\left[
u_{R,k}^{(N)}\left(  t\right)  \right]  ^{r}\left[  u_{W,k}^{(N)}\left(
t\right)  \right]  ^{d_{1}-m-r}. \label{Equ-6-1}%
\end{align}

The following theorem simplifies expressions for the probabilities
$W_{W,k}^{\left(  N\right)  }(u_{W,k-1},u_{W,k}$; $u_{R,k-1},u_{R,k};t)$ and
$W_{R,k}^{\left(  N\right)  }\left(  u_{W,k-1},u_{W,k};u_{R,k-1}%
,u_{R,k};t\right)  $ for $k\geq2$, while its proof is similar to that in
Theorem 1 of Li et al. \cite{Li:2014a} and is omitted here. Note that the
simplified expressions will be a key in our later study, for example, the
system of mean-field equations can be simplified significantly and the fixed
point can be computed effectively.

\begin{The}
\label{The:Inv}%
\[
W_{W,1}^{\left(  N\right)  }\left(  u_{W,0},u_{W,1};u_{R,1};t\right)
=\sum_{m=1}^{d_{1}}C_{d_{1}}^{m}\left[  u_{W,0}^{(N)}\left(  t\right)
-u_{W,1}^{(N)}\left(  t\right)  \right]  ^{m-1}\left[  u_{W,1}^{(N)}%
(t)+u_{R.1}^{(N)}\left(  t\right)  \right]  ^{d_{1}-m},
\]
and for $k\geq2$%
\begin{align*}
&  W_{W,k}^{\left(  N\right)  }\left(  u_{W,k-1},u_{W,k};u_{R,k-1}%
,u_{R,k};t\right)  =W_{R,k}^{\left(  N\right)  }\left(  u_{W,k-1}%
,u_{W,k};u_{R,k-1},u_{R,k};t\right) \\
&  =\sum_{m=1}^{d_{1}}C_{d_{1}}^{m}\left[  u_{W,k-1}^{(N)}\left(  t\right)
-u_{W,k}^{(N)}\left(  t\right)  +u_{R,k-1}^{(N)}\left(  t\right)
-u_{R.k}^{(N)}\left(  t\right)  \right]  ^{m-1}\left[  u_{W,k}^{(N)}%
(t)+u_{R.k}^{(N)}\left(  t\right)  \right]  ^{d_{1}-m}.
\end{align*}
\end{The}

Using Theorem \ref{The:Inv}, we set%
\[
L_{1}^{\left(  N\right)  }\left(  u_{W,0},u_{W,1};u_{R,1};t\right)
=W_{W,1}^{\left(  N\right)  }\left(  u_{W,0},u_{W,1};u_{R,1};t\right)
\]
and for $k\geq2$%
\begin{align*}
L_{k}^{\left(  N\right)  }\left(  u_{W,k-1},u_{W,k};u_{R,k-1},u_{R,k}%
;t\right)   &  =W_{W,k}^{\left(  N\right)  }\left(  u_{W,k-1},u_{W,k}%
;u_{R,k-1},u_{R,k};t\right) \\
&  =W_{R,k}^{\left(  N\right)  }\left(  u_{W,k-1},u_{W,k};u_{R,k-1}%
,u_{R,k};t\right)  .
\end{align*}

\textbf{(A.}$2$\textbf{): Observing Both the Queue Length and the States of
the Chosen Servers}

In this case, the arriving customer has a priority for joining one working
server with the shortest queue length.\textbf{ }Upon arrival, each customer
chooses $d_{1}\geq1$ servers from the $N$ servers independently and uniformly
at random, and joins the one whose queue length is the shortest among the
$d_{1}$ servers. If the servers with the shortest queue length contain at
least one working server and at least one server in repair, then the arriving
customer must randomly join one of the working servers with the shortest queue
length. If there is a tie, the working servers with the shortest queue length
are chosen randomly.

It is seen that the only difference from \textbf{(A.}$1$\textbf{)} is that the
arriving customer can not join one of the repairing servers with the shortest
queue length when there exists at least one working server with the shortest
queue length. Based on this, we have

\textbf{a) }The probabilities $W_{W,1}^{\left(  N\right)  }\left(
u_{W,0},u_{W,1};u_{R,1};t\right)  $ and $W_{W,k}^{\left(  N\right)  }\left(
u_{W,k-1},u_{W,k};u_{R,k-1},u_{R,k};t\right)  $ for $k\geq2$ are the same as
those in \textbf{(A.}$1$\textbf{)}.

\textbf{b)} Comparing with the probabilities $W_{R,k}^{\left(  N\right)
}\left(  u_{W,k-1},u_{W,k};u_{R,k-1},u_{R,k};t\right)  $ in \textbf{(A.}%
$1$\textbf{)}, for \textbf{(A.}$2$\textbf{)}\ we obtain that for $k\geq2$%
\begin{align*}
&  \mathcal{W}_{R,k}^{\left(  N\right)  }\left(  u_{W,k-1},u_{W,k}%
;u_{R,k-1},u_{R,k};t\right)  =\sum_{m=1}^{d_{1}}C_{d_{1}}^{m}\left[
u_{R,k-1}^{(N)}\left(  t\right)  -u_{R.k}^{(N)}\left(  t\right)  \right]
^{m-1}\left[  u_{R,k}^{(N)}\left(  t\right)  \right]  ^{d_{1}-m}\\
&  +\sum_{m=1}^{d_{1}-1}C_{d_{1}}^{m}\left[  u_{R,k-1}^{(N)}\left(  t\right)
-u_{R.k}^{(N)}\left(  t\right)  \right]  ^{m-1}\sum_{j=1}^{d_{1}-m}C_{d_{1}%
-m}^{j}\left[  u_{R,k}^{(N)}\left(  t\right)  \right]  ^{d_{1}-m-j}\left[
u_{W,k}^{(N)}\left(  t\right)  \right]  ^{j}.
\end{align*}
Note that Part III of computing $W_{R,k}^{\left(  N\right)  }\left(
u_{W,k-1},u_{W,k};u_{R,k-1},u_{R,k};t\right)  $ in \textbf{(A.}$1$\textbf{)}
is omitted by utilizing the information of \textbf{(A.}$2$\textbf{)}.

\subsection{The repair processes}

Now we provide the probability representations for the repair processes in two
cases: (R.$1$) and (R.$2$).

\textbf{(R.}$1$\textbf{): Each Server Has One Repairman}

In this case, there are $N$ repairmen corresponding to the $N$ servers, hence
each server has one repairman. Since the repair time is exponentially
distributed with repair rate $\beta$, it is seen from Li et al. \cite{Li:2006}
if the service time of each server is of phase type with irreducible matrix
representation $\left(  \tau,T\right)  $, where%
\[
\tau=\left(  1,0\right)  ,\text{ \ }T=\left(
\begin{array}
[c]{cc}%
-\left(  \mu+\alpha\right)  & \alpha\\
\beta & -\beta
\end{array}
\right)  ,
\]
then the repairable supermarket model is equivalent to a supermarket model
with Poisson inputs and PH service times, as discussed in Li and Lui
\cite{Li:2014}.

\textbf{(R.}$2$\textbf{): A Super Repairman}

In this case, there is a single super repairman whose repair time is
exponentially distributed with repair rate $N\beta$. The repairman chooses
$d_{2}$ servers from the $N$ servers independently and uniformly at random. If
all the selected $d_{2}$ servers are in working condition, the repairman is
idle; if at least one of the selected $d_{2}$ servers is failed, then the
repairman attends one failed server with the longest queue. If there is a tie,
the repairman select a server randomly.

The rate that the repairman randomly chooses one of the failed servers with
the longest queue length $k$ and the queue lengths of the other $d_{2}-1$
selected servers are not longer than $k$ is given by%
\[
N\beta\sum_{m=1}^{d_{2}}C_{d_{2}}^{m}\left[  u_{R,1}^{\left(  N\right)
}\left(  t\right)  \right]  ^{m}\left[  u_{W,0}^{\left(  N\right)  }\left(
t\right)  -u_{W,2}^{\left(  N\right)  }\left(  t\right)  \right]  ^{d_{2}%
-m}\text{d}t\overset{\text{def}}{=}N\beta I_{1}\left(  u_{R,1}^{\left(
N\right)  },u_{W,0}^{\left(  N\right)  },u_{W,2}^{\left(  N\right)
};t\right)  \text{d}t,
\]
and for $k\geq2$%
\begin{align*}
&  N\beta\sum_{m=1}^{d_{2}}C_{d_{2}}^{m}\left\{  \sum_{m_{1}=1}^{m}%
C_{m}^{m_{1}}\left[  u_{R,k}^{\left(  N\right)  }\left(  t\right)  \right]
^{m_{1}}\left[  u_{R,1}^{\left(  N\right)  }\left(  t\right)  -u_{R,k}%
^{\left(  N\right)  }\left(  t\right)  \right]  ^{m-m_{1}}\right\}  \left[
u_{W,0}^{\left(  N\right)  }\left(  t\right)  -u_{W,k+1}^{\left(  N\right)
}\left(  t\right)  \right]  ^{d_{2}-m}\text{d}t\\
&  \overset{\text{def}}{=}N\beta I_{k}\left(  u_{R,1}^{\left(  N\right)
},u_{R,k}^{\left(  N\right)  },u_{W,0}^{\left(  N\right)  },u_{W,k+1}^{\left(
N\right)  };t\right)  \text{d}t.
\end{align*}
Using $u_{W,0}^{\left(  N\right)  }\left(  t\right)  +u_{R,1}^{\left(
N\right)  }\left(  t\right)  =1$, we can further simplify%
\begin{align}
I_{1}\left(  u_{R,1}^{\left(  N\right)  },u_{W,0}^{\left(  N\right)  }%
,u_{W,2}^{\left(  N\right)  };t\right)   &  =\left\{  \left[  1-u_{W,2}%
^{\left(  N\right)  }\left(  t\right)  \right]  ^{d_{2}}-\left[
u_{W,0}^{\left(  N\right)  }\left(  t\right)  -u_{W,2}^{\left(  N\right)
}\left(  t\right)  \right]  ^{d_{2}}\right\} \nonumber\\
&  =\left[  1-u_{W,2}^{\left(  N\right)  }\left(  t\right)  \right]  ^{d_{2}%
}-\left[  1-u_{R,1}^{\left(  N\right)  }\left(  t\right)  -u_{W,2}^{\left(
N\right)  }\left(  t\right)  \right]  ^{d_{2}}\nonumber\\
&  \overset{\text{def}}{=}I_{1}\left(  u_{R,1}^{\left(  N\right)  }\left(
t\right)  ,u_{W,2}^{\left(  N\right)  };t\right)  \label{I-1}%
\end{align}
and for $k\geq2$%
\begin{align}
I_{k}\left(  u_{R,1}^{\left(  N\right)  },u_{R,k}^{\left(  N\right)  }%
,u_{W,0}^{\left(  N\right)  },u_{W,k+1}^{\left(  N\right)  };t\right)   &
=\left[  1-u_{W,k+1}^{\left(  N\right)  }\left(  t\right)  \right]  ^{d_{2}%
}-\left[  1-u_{R,k}^{\left(  N\right)  }\left(  t\right)  -u_{W,k+1}^{\left(
N\right)  }\left(  t\right)  \right]  ^{d_{2}}\nonumber\\
&  \overset{\text{def}}{=}I_{k}\left(  u_{R,k}^{\left(  N\right)  }%
,u_{W,k+1}^{\left(  N\right)  };t\right)  . \label{I-k}%
\end{align}

\section{The Mean-Field Equations}

In this section, for each of the four interrelated supermarket models with
repairable servers, we set up an infinite-dimensional system of mean-field
equations. To this end, we present a detailed analysis only for the first
model, while the other three models can be simply discussed on a similar line.

\subsection{Model I ((A.$1$)\textbf{ }and (R.$1$))}

For (A.$1$)\textbf{ }and (R.$1$), the probabilities $W_{W,1}^{\left(
N\right)  }\left(  u_{W,0},u_{W,1};u_{R,1};t\right)  $, $W_{W,k}^{\left(
N\right)  }\left(  u_{W,k-1},u_{W,k};u_{R,k-1},u_{R,k};t\right)  $ and
$W_{R,k}^{\left(  N\right)  }\left(  u_{W,k-1},u_{W,k};u_{R,k-1}%
,u_{R,k};t\right)  $ for $k\geq2$ are given in (\ref{Equ-4-1}), (\ref{Equ-4-2}%
) and (\ref{Equ-6-1}), which are further simplified in Theorem \ref{The:Inv}.

Now, we consider the service and repair processes. The rate that a customer
leaves one server queued by $k$ customers is given by%
\begin{equation}
N\mu\left[  u_{W,k}^{(N)}(t)-u_{W,k+1}^{(N)}(t)\right]  \text{d}t.
\label{Equ4-1}%
\end{equation}
The rate that one working server with at least $k$ customers fails is given by%
\begin{equation}
N\alpha u_{W,k}^{(N)}(t)\text{d}t. \label{Equ4-2}%
\end{equation}
The rate that one failed server with at least $k$ customers is repaired is
given by%
\begin{equation}
N\beta u_{R,k}^{\left(  N\right)  }\left(  t\right)  \text{d}t. \label{Equ4-3}%
\end{equation}
Based on Equation (\ref{Equ4}), and Equations (\ref{Equ4-1}) to (\ref{Equ4-3}%
), we obtain%
\begin{align}
\frac{\text{d}}{\text{d}t}u_{W,k}^{(N)}\left(  t\right)  =  &  \lambda\left[
u_{W,k-1}^{(N)}\left(  t\right)  -u_{W,k}^{(N)}\left(  t\right)  \right]
W_{W,k}\left(  u_{W,k-1},u_{W,k};u_{R,k-1},u_{R,k};t\right) \nonumber\\
&  -\mu\left[  u_{W,k}^{(N)}(t)-u_{W,k+1}^{(N)}(t)\right]  -\alpha
u_{W,k}^{(N)}(t)+\beta u_{R,k}^{\left(  N\right)  }\left(  t\right)  .
\label{Equ5}%
\end{align}
In addition, it follows from (\ref{Equ6}) that%
\begin{align}
\frac{\text{d}}{\text{d}t}u_{R,k}^{(N)}\left(  t\right)  =  &  \lambda\left[
u_{R,k-1}^{(N)}\left(  t\right)  -u_{R.k}^{(N)}\left(  t\right)  \right]
W_{R,k}^{\left(  N\right)  }\left(  u_{W,k-1},u_{W,k};u_{R,k-1},u_{R,k}%
;t\right) \nonumber\\
&  +\alpha u_{W,k}^{(N)}(t)-\beta u_{R,k}^{(N)}\left(  t\right)  .
\label{Equ7}%
\end{align}

Based on the similar analysis to (\ref{Equ5}) and (\ref{Equ7}), we can set up
an infinite-dimensional system of mean-field equations satisfied by the
expected fraction vector $\mathbf{u}^{(N)}(t)=\left(  \mathbf{u}_{W}%
^{(N)}(t),\mathbf{u}_{R}^{(N)}(t)\right)  $ as follows:%
\begin{equation}
\frac{\text{d}}{\text{d}t}u_{W,0}^{(N)}(t)=-\alpha u_{W,1}^{(N)}(t)+\beta
u_{R,1}^{\left(  N\right)  }\left(  t\right)  , \label{Equap8}%
\end{equation}%
\begin{align}
\frac{\text{d}}{\text{d}t}u_{W,1}^{(N)}(t)=  &  \lambda\left[  u_{W,0}%
^{(N)}\left(  t\right)  -u_{W,1}^{(N)}\left(  t\right)  \right]
W_{W,1}^{\left(  N\right)  }\left(  u_{W,0},u_{W,1};u_{R,1};t\right)
\nonumber\\
&  -\mu\left[  u_{W,1}^{(N)}(t)-u_{W,2}^{(N)}(t)\right]  -\alpha u_{W,1}%
^{(N)}(t)+\beta u_{R,1}^{\left(  N\right)  }\left(  t\right)  ,
\label{Equap8-1}%
\end{align}
for $k\geq2$%
\begin{align}
\frac{\text{d}}{\text{d}t}u_{W,k}^{(N)}\left(  t\right)  =  &  \lambda\left[
u_{W,k-1}^{(N)}\left(  t\right)  -u_{W,k}^{(N)}\left(  t\right)  \right]
W_{W,k}^{\left(  N\right)  }\left(  u_{W,k-1},u_{W,k};u_{R,k-1},u_{R,k}%
;t\right) \nonumber\\
&  -\mu\left[  u_{W,k}^{(N)}(t)-u_{W,k+1}^{(N)}(t)\right]  -\alpha
u_{W,k}^{(N)}(t)+\beta u_{R,k}^{\left(  N\right)  }\left(  t\right)
\label{Equap9}%
\end{align}
and%
\begin{align}
\frac{\text{d}}{\text{d}t}u_{R,k}^{(N)}\left(  t\right)  =  &  \lambda\left[
u_{R,k-1}^{(N)}\left(  t\right)  -u_{R.k}^{(N)}\left(  t\right)  \right]
W_{R,k}^{\left(  N\right)  }\left(  u_{W,k-1},u_{W,k};u_{R,k-1},u_{R,k}%
;t\right) \nonumber\\
&  +\alpha u_{W,k}^{(N)}(t)-\beta u_{R,k}^{\left(  N\right)  }\left(
t\right)  , \label{Equap10}%
\end{align}
with the boundary condition%
\begin{equation}
u_{W,0}^{(N)}\left(  t\right)  +u_{R,1}^{(N)}\left(  t\right)  =1,\text{
\ \ }t\geq0, \label{Equap12}%
\end{equation}
and the initial conditions%
\begin{equation}
\left\{
\begin{array}
[c]{cc}%
u_{W,k}^{(N)}\left(  0\right)  =g_{k}, & \text{\ }k\geq0,\\
u_{R,l}^{(N)}\left(  0\right)  =h_{l}, & l\geq1.
\end{array}
\right.  \text{ \ } \label{Equap13}%
\end{equation}
where%
\[
1\geq g_{0}\geq g_{1}\geq g_{2}\geq\cdots\geq0,
\]%
\[
1\geq h_{1}\geq h_{2}\geq h_{3}\geq\cdots\geq0,
\]
with%
\[
g_{0}+h_{1}=1.
\]

It follows from Theorem \ref{The:Inv} that%
\[
L_{1}^{\left(  N\right)  }\left(  u_{W,0},u_{W,1};u_{R,1};t\right)
=W_{W,1}^{\left(  N\right)  }\left(  u_{W,0},u_{W,1};u_{R,1};t\right)
\]
and for $k\geq2$%
\begin{align}
L_{k}^{\left(  N\right)  }\left(  u_{W,k-1},u_{W,k};u_{R,k-1},u_{R,k}%
;t\right)   &  =W_{W,k}^{\left(  N\right)  }\left(  u_{W,k-1},u_{W,k}%
;u_{R,k-1},u_{R,k};t\right) \nonumber\\
&  =W_{R,k}^{\left(  N\right)  }\left(  u_{W,k-1},u_{W,k};u_{R,k-1}%
,u_{R,k};t\right)  . \label{Equ7-1}%
\end{align}
Using $L_{1}^{\left(  N\right)  }\left(  u_{W,0},u_{W,1};u_{R,1};t\right)  $
and $L_{k}^{\left(  N\right)  }\left(  u_{W,k-1},u_{W,k};u_{R,k-1}%
,u_{R,k};t\right)  $ for $k\geq2$, Equations (\ref{Equap8}) to (\ref{Equap13})
can further be simplified as%
\begin{equation}
\frac{\text{d}}{\text{d}t}u_{W,0}^{(N)}(t)=-\alpha u_{W,1}^{(N)}(t)+\beta
u_{R,1}^{\left(  N\right)  }\left(  t\right)  , \label{Equ8}%
\end{equation}%
\begin{align}
\frac{\text{d}}{\text{d}t}u_{W,1}^{(N)}(t)=  &  \lambda\left[  u_{W,0}%
^{(N)}\left(  t\right)  -u_{W,1}^{(N)}\left(  t\right)  \right]
L_{1}^{\left(  N\right)  }\left(  u_{W,0},u_{W,1};u_{R,1};t\right) \nonumber\\
&  -\mu\left[  u_{W,1}^{(N)}(t)-u_{W,2}^{(N)}(t)\right]  -\alpha u_{W,1}%
^{(N)}(t)+\beta u_{R,1}^{\left(  N\right)  }\left(  t\right)  , \label{Equ8-1}%
\end{align}
for $k\geq2$%
\begin{align}
\frac{\text{d}}{\text{d}t}u_{W,k}^{(N)}\left(  t\right)  =  &  \lambda\left[
u_{W,k-1}^{(N)}\left(  t\right)  -u_{W,k}^{(N)}\left(  t\right)  \right]
L_{k}^{\left(  N\right)  }\left(  u_{W,k-1},u_{W,k};u_{R,k-1},u_{R,k};t\right)
\nonumber\\
&  -\mu\left[  u_{W,k}^{(N)}(t)-u_{W,k+1}^{(N)}(t)\right]  -\alpha
u_{W,k}^{(N)}(t)+\beta u_{R,k}^{\left(  N\right)  }\left(  t\right)
\label{Equ9}%
\end{align}
and%
\begin{align}
\frac{\text{d}}{\text{d}t}u_{R,k}^{(N)}\left(  t\right)  =  &  \lambda\left[
u_{R,k-1}^{(N)}\left(  t\right)  -u_{R.k}^{(N)}\left(  t\right)  \right]
L_{k}^{\left(  N\right)  }\left(  u_{W,k-1},u_{W,k};u_{R,k-1},u_{R,k};t\right)
\nonumber\\
&  +\alpha u_{W,k}^{(N)}(t)-\beta u_{R,k}^{\left(  N\right)  }\left(
t\right)  , \label{Equ10}%
\end{align}
with the boundary condition%
\begin{equation}
u_{W,0}^{(N)}\left(  t\right)  +u_{R,1}^{(N)}\left(  t\right)  =1,\text{
\ \ }t\geq0, \label{Equ12}%
\end{equation}
and the initial conditions%
\begin{equation}
\left\{
\begin{array}
[c]{cc}%
u_{W,k}^{(N)}\left(  0\right)  =g_{k}, & \text{\ }k\geq0,\\
u_{R,l}^{(N)}\left(  0\right)  =h_{l}, & l\geq1.
\end{array}
\right.  \text{ \ } \label{Equ13}%
\end{equation}

\subsection{Model II ((A.$1$)\textbf{ }and (R.$2$))}

In this model, for (R.$2$) it follows from (\ref{I-1}) and (\ref{I-k}) that
for $k\geq1$%
\[
I_{k}\left(  u_{R,k}^{\left(  N\right)  },u_{W,k+1}^{\left(  N\right)
};t\right)  =\left[  1-u_{W,k+1}^{\left(  N\right)  }\left(  t\right)
\right]  ^{d_{2}}-\left[  1-u_{R,k}^{\left(  N\right)  }\left(  t\right)
-u_{W,k+1}^{\left(  N\right)  }\left(  t\right)  \right]  ^{d_{2}}.
\]
Hence the dynamic routine selection scheme (R.$2$) shows that for $k\geq1$,
$\beta I_{k}\left(  u_{R,k}^{\left(  N\right)  },u_{W,k+1}^{\left(  N\right)
};t\right)  $ will take the place of $\beta u_{R,k}^{\left(  N\right)
}\left(  t\right)  $ in the systems of mean-field equations (\ref{Equ8}) to
(\ref{Equ13}). Based on this, we obtain%
\begin{equation}
\frac{\text{d}}{\text{d}t}u_{W,0}^{(N)}(t)=-\alpha u_{W,1}^{(N)}(t)+\beta
I_{1}\left(  u_{R,1}^{\left(  N\right)  }\left(  t\right)  ,u_{W,2}^{\left(
N\right)  };t\right)  , \label{MII-1}%
\end{equation}%
\begin{align}
\frac{\text{d}}{\text{d}t}u_{W,1}^{(N)}(t)=  &  \lambda\left[  u_{W,0}%
^{(N)}\left(  t\right)  -u_{W,1}^{(N)}\left(  t\right)  \right]
L_{1}^{\left(  N\right)  }\left(  u_{W,0},u_{W,1};u_{R,1};t\right) \nonumber\\
&  -\mu\left[  u_{W,1}^{(N)}(t)-u_{W,2}^{(N)}(t)\right]  -\alpha u_{W,1}%
^{(N)}(t)+\beta I_{1}\left(  u_{R,1}^{\left(  N\right)  }\left(  t\right)
,u_{W,2}^{\left(  N\right)  };t\right)  , \label{MII-2}%
\end{align}
for $k\geq2$%
\begin{align}
\frac{\text{d}}{\text{d}t}u_{W,k}^{(N)}\left(  t\right)  =  &  \lambda\left[
u_{W,k-1}^{(N)}\left(  t\right)  -u_{W,k}^{(N)}\left(  t\right)  \right]
L_{k}^{\left(  N\right)  }\left(  u_{W,k-1},u_{W,k};u_{R,k-1},u_{R,k};t\right)
\nonumber\\
&  -\mu\left[  u_{W,k}^{(N)}(t)-u_{W,k+1}^{(N)}(t)\right]  -\alpha
u_{W,k}^{(N)}(t)+\beta I_{k}\left(  u_{R,k}^{\left(  N\right)  }%
,u_{W,k+1}^{\left(  N\right)  };t\right)  \label{MII-3}%
\end{align}
and%
\begin{align}
\frac{\text{d}}{\text{d}t}u_{R,k}^{(N)}\left(  t\right)  =  &  \lambda\left[
u_{R,k-1}^{(N)}\left(  t\right)  -u_{R.k}^{(N)}\left(  t\right)  \right]
L_{k}^{\left(  N\right)  }\left(  u_{W,k-1},u_{W,k};u_{R,k-1},u_{R,k};t\right)
\nonumber\\
&  +\alpha u_{W,k}^{(N)}(t)-\beta I_{k}\left(  u_{R,k}^{\left(  N\right)
},u_{W,k+1}^{\left(  N\right)  };t\right)  , \label{MII-4}%
\end{align}
with the boundary condition%
\begin{equation}
u_{W,0}^{(N)}\left(  t\right)  +u_{R,1}^{(N)}\left(  t\right)  =1,\text{
\ \ }t\geq0, \label{MII-5}%
\end{equation}
and the initial conditions%
\begin{equation}
\left\{
\begin{array}
[c]{cc}%
u_{W,k}^{(N)}\left(  0\right)  =g_{k}, & \text{\ }k\geq0,\\
u_{R,l}^{(N)}\left(  0\right)  =h_{l}, & l\geq1.
\end{array}
\right.  \text{ \ } \label{MII-6}%
\end{equation}

\subsection{Model III ((A.$2$)\textbf{ }and (R.$1$))}

In this model, the only difference is that an arriving customer cannot join
the server in repair with the shortest queue length when there exists at least
one working server with the shortest queue length. Thus we obtain%
\begin{align}
&  \mathcal{W}_{R,k}^{\left(  N\right)  }\left(  u_{W,k-1},u_{W,k}%
;u_{R,k-1},u_{R,k};t\right)  =\sum_{m=1}^{d_{1}}C_{d_{1}}^{m}\left[
u_{R,k-1}^{(N)}\left(  t\right)  -u_{R.k}^{(N)}\left(  t\right)  \right]
^{m-1}\left[  u_{R,k}^{(N)}\left(  t\right)  \right]  ^{d_{1}-m}\nonumber\\
&  +\sum_{m=1}^{d_{1}-1}C_{d_{1}}^{m}\left[  u_{R,k-1}^{(N)}\left(  t\right)
-u_{R.k}^{(N)}\left(  t\right)  \right]  ^{m-1}\sum_{j=1}^{d_{1}-m}C_{d_{1}%
-m}^{j}\left[  u_{R,k}^{(N)}\left(  t\right)  \right]  ^{d_{1}-m-j}\left[
u_{W,k}^{(N)}\left(  t\right)  \right]  ^{j} \label{Equa-1}%
\end{align}
and%
\begin{align*}
&  \left[  u_{R,k-1}^{(N)}\left(  t\right)  -u_{R.k}^{(N)}\left(  t\right)
\right]  \mathcal{W}_{R,k}^{\left(  N\right)  }\left(  u_{W,k-1}%
,u_{W,k};u_{R,k-1},u_{R,k};t\right) \\
&  =\left[  u_{R,k-1}^{\left(  N\right)  }\left(  t\right)  +u_{W,k}^{\left(
N\right)  }\left(  t\right)  \right]  ^{d_{1}}-\left[  u_{W,k}^{\left(
N\right)  }\left(  t\right)  +u_{R,k}^{\left(  N\right)  }\left(  t\right)
\right]  ^{d_{1}}.
\end{align*}
It is easy to see from (\ref{Equ-6-1}), (\ref{Equa-1}) and Theorem
\ref{The:Inv} that%
\begin{equation}
\mathcal{W}_{R,k}^{\left(  N\right)  }\left(  u_{W,k-1},u_{W,k};u_{R,k-1}%
,u_{R,k};t\right)  \neq L_{k}^{\left(  N\right)  }\left(  u_{W,k-1}%
,u_{W,k};u_{R,k-1},u_{R,k};t\right)  . \label{Equa-2}%
\end{equation}
Thus (A.$2$) indicates that $\mathcal{W}_{R,k}^{\left(  N\right)  }\left(
u_{W,k-1},u_{W,k};u_{R,k-1},u_{R,k};t\right)  $ needs to replace
$W_{R,k}^{\left(  N\right)  }(u_{W,k-1}$, $u_{W,k};u_{R,k-1},u_{R,k};t)$ in
the systems of mean-field equations (\ref{Equ8}) to (\ref{Equ13}).

On the other hand, except of (\ref{Equa-2}), we still have%
\[
W_{W,1}^{\left(  N\right)  }\left(  u_{W,0},u_{W,1};u_{R,1};t\right)
=L_{1}^{\left(  N\right)  }\left(  u_{W,0},u_{W,1};u_{R,1};t\right)
\]
and for $k\geq2$%
\[
W_{W,k}^{\left(  N\right)  }\left(  u_{W,k-1},u_{W,k};u_{R,k-1},u_{R,k}%
;t\right)  =L_{k}^{\left(  N\right)  }\left(  u_{W,k-1},u_{W,k};u_{R,k-1}%
,u_{R,k};t\right)  .
\]

A similar analysis to the systems of mean-field equations (\ref{Equ8}) to
(\ref{Equ13}), we obtain%
\begin{equation}
\frac{\text{d}}{\text{d}t}u_{W,0}^{(N)}(t)=-\alpha u_{W,1}^{(N)}(t)+\beta
u_{R,1}^{\left(  N\right)  }\left(  t\right)  , \label{MIII-1}%
\end{equation}%
\begin{align}
\frac{\text{d}}{\text{d}t}u_{W,1}^{(N)}(t)=  &  \lambda\left[  u_{W,0}%
^{(N)}\left(  t\right)  -u_{W,1}^{(N)}\left(  t\right)  \right]
L_{1}^{\left(  N\right)  }\left(  u_{W,0},u_{W,1};u_{R,1};t\right) \nonumber\\
&  -\mu\left[  u_{W,1}^{(N)}(t)-u_{W,2}^{(N)}(t)\right]  -\alpha u_{W,1}%
^{(N)}(t)+\beta u_{R,1}^{\left(  N\right)  }\left(  t\right)  , \label{MIII-2}%
\end{align}
for $k\geq2$%
\begin{align}
\frac{\text{d}}{\text{d}t}u_{W,k}^{(N)}\left(  t\right)  =  &  \lambda\left[
u_{W,k-1}^{(N)}\left(  t\right)  -u_{W,k}^{(N)}\left(  t\right)  \right]
L_{k}^{\left(  N\right)  }\left(  u_{W,k-1},u_{W,k};u_{R,k-1},u_{R,k};t\right)
\nonumber\\
&  -\mu\left[  u_{W,k}^{(N)}(t)-u_{W,k+1}^{(N)}(t)\right]  -\alpha
u_{W,k}^{(N)}(t)+\beta u_{R,k}^{\left(  N\right)  }\left(  t\right)
\label{MIII-3}%
\end{align}
and%
\begin{align}
\frac{\text{d}}{\text{d}t}u_{R,k}^{(N)}\left(  t\right)  =  &  \lambda\left[
u_{R,k-1}^{(N)}\left(  t\right)  -u_{R.k}^{(N)}\left(  t\right)  \right]
\mathcal{W}_{R,k}^{\left(  N\right)  }\left(  u_{W,k-1},u_{W,k};u_{R,k-1}%
,u_{R,k};t\right) \nonumber\\
&  +\alpha u_{W,k}^{(N)}(t)-\beta u_{R,k}^{\left(  N\right)  }\left(
t\right)  , \label{MIII-4}%
\end{align}
with the boundary condition%
\begin{equation}
u_{W,0}^{(N)}\left(  t\right)  +u_{R,1}^{(N)}\left(  t\right)  =1,\text{
\ \ }t\geq0, \label{MIII-5}%
\end{equation}
and the initial conditions%
\begin{equation}
\left\{
\begin{array}
[c]{cc}%
u_{W,k}^{(N)}\left(  0\right)  =g_{k}, & \text{\ }k\geq0,\\
u_{R,l}^{(N)}\left(  0\right)  =h_{l}, & l\geq1.
\end{array}
\right.  \text{ \ } \label{MIII-6}%
\end{equation}

\subsection{Model IV ((A.$2$)\textbf{ }and (R.$2$))}

Since (A.$2$) needs $\mathcal{W}_{R,k}^{\left(  N\right)  }\left(
u_{W,k-1},u_{W,k};u_{R,k-1},u_{R,k};t\right)  $ replacing $W_{R,k}^{\left(
N\right)  }\left(  u_{W,k-1},u_{W,k};u_{R,k-1},u_{R,k};t\right)  $, and
(R.$2$) needs $\beta I_{k}\left(  u_{R,k}^{\left(  N\right)  },u_{W,k+1}%
^{\left(  N\right)  };t\right)  $ taking the place of $\beta u_{R,k}^{\left(
N\right)  }\left(  t\right)  $. Thus we obtain%
\begin{equation}
\frac{\text{d}}{\text{d}t}u_{W,0}^{(N)}(t)=-\alpha u_{W,1}^{(N)}(t)+\beta
I_{1}\left(  u_{R,1}^{\left(  N\right)  }\left(  t\right)  ,u_{W,2}^{\left(
N\right)  };t\right)  , \label{MIV-1}%
\end{equation}%
\begin{align}
\frac{\text{d}}{\text{d}t}u_{W,1}^{(N)}(t)=  &  \lambda\left[  u_{W,0}%
^{(N)}\left(  t\right)  -u_{W,1}^{(N)}\left(  t\right)  \right]
L_{1}^{\left(  N\right)  }\left(  u_{W,0},u_{W,1};u_{R,1};t\right) \nonumber\\
&  -\mu\left[  u_{W,1}^{(N)}(t)-u_{W,2}^{(N)}(t)\right]  -\alpha u_{W,1}%
^{(N)}(t)+\beta I_{1}\left(  u_{R,1}^{\left(  N\right)  }\left(  t\right)
,u_{W,2}^{\left(  N\right)  };t\right)  , \label{MIV-2}%
\end{align}
for $k\geq2$%
\begin{align}
\frac{\text{d}}{\text{d}t}u_{W,k}^{(N)}\left(  t\right)  =  &  \lambda\left[
u_{W,k-1}^{(N)}\left(  t\right)  -u_{W,k}^{(N)}\left(  t\right)  \right]
L_{k}^{\left(  N\right)  }\left(  u_{W,k-1},u_{W,k};u_{R,k-1},u_{R,k};t\right)
\nonumber\\
&  -\mu\left[  u_{W,k}^{(N)}(t)-u_{W,k+1}^{(N)}(t)\right]  -\alpha
u_{W,k}^{(N)}(t)+\beta I_{k}\left(  u_{R,k}^{\left(  N\right)  }%
,u_{W,k+1}^{\left(  N\right)  };t\right)  \label{MIV-3}%
\end{align}
and%
\begin{align}
\frac{\text{d}}{\text{d}t}u_{R,k}^{(N)}\left(  t\right)  =  &  \lambda\left[
u_{R,k-1}^{(N)}\left(  t\right)  -u_{R.k}^{(N)}\left(  t\right)  \right]
\mathcal{W}_{R,k}^{\left(  N\right)  }\left(  u_{W,k-1},u_{W,k};u_{R,k-1}%
,u_{R,k};t\right) \nonumber\\
&  +\alpha u_{W,k}^{(N)}(t)-\beta I_{k}\left(  u_{R,k}^{\left(  N\right)
},u_{W,k+1}^{\left(  N\right)  };t\right)  , \label{MIV-4}%
\end{align}
with the boundary condition%
\begin{equation}
u_{W,0}^{(N)}\left(  t\right)  +u_{R,1}^{(N)}\left(  t\right)  =1,\text{
\ \ }t\geq0, \label{MIV-5}%
\end{equation}
and the initial conditions%
\begin{equation}
\left\{
\begin{array}
[c]{cc}%
u_{W,k}^{(N)}\left(  0\right)  =g_{k}, & \text{\ }k\geq0,\\
u_{R,l}^{(N)}\left(  0\right)  =h_{l}, & l\geq1.
\end{array}
\right.  \text{ \ } \label{MIV-6}%
\end{equation}

\begin{Rem}
From the four systems of mean-field equations, we find that to set up the
systems of mean-field equations, two key rules must be followed as follows:

(1) If (A.$1$) $\rightarrow$ (A.$2$), then $\mathcal{W}_{R,k}^{\left(
N\right)  }\left(  u_{W,k-1},u_{W,k};u_{R,k-1},u_{R,k};t\right)  $ takes the
place of $W_{R,k}^{\left(  N\right)  }(u_{W,k-1}$, $u_{W,k};u_{R,k-1}%
,u_{R,k};t)$, and

(2) if (R.$1$) $\rightarrow$ (R.$2$), then $\beta I_{k}\left(  u_{R,k}%
^{\left(  N\right)  },u_{W,k+1}^{\left(  N\right)  };t\right)  $ replaces
$\beta u_{R,k}^{\left(  N\right)  }\left(  t\right)  $.
\end{Rem}

\section{The Fixed Point}

In this section, we discuss the fixed points for the systems of mean-field
equations, and show that the fixed points can be determined by the systems of
nonlinear equations. Specifically, we indicate that the nonlinear structure
makes the analytical solution of the fixed points too complicated and even
impossible. Since such a fixed point plays a key role in performance analysis
of the supermarket models with repairable servers, it is interesting to
develop numerical computation in the study of complex supermarket models.

\subsection{A double limit}

We discuss a double limit of the expected fraction vector function
$\mathbf{u}^{\left(  N\right)  }\left(  t\right)  $ as $N\rightarrow\infty$
and $t\rightarrow+\infty$.

The following lemma provides a sufficient condition under which each of the
four interrelated supermarket models with $N$ identical repairable servers is stable.

\begin{Lem}
Each of the four supermarket model with $N$ identical and repairable servers
and two choice numbers $d_{1},d_{2}\geq1$ is stable if $\widetilde{\rho}%
=\rho\left(  1+\alpha/\beta\right)  <1$, where $\rho=\lambda/\mu$.
\end{Lem}

\textbf{Proof:} \ If $d_{1}=d_{2}=1$, then each of the four supermarket models
of $N$ identical repairable servers is equivalent to a system of $N$
independent M/M/1 queues with repairable servers. From Li et al.
\cite{Li:2006}, it is easy to see that such a repairable M/M/1 queue is stable
if $\widetilde{\rho}<1$. Using a coupling method, as given in Theorems 4 and 5
of Martin and Suhov \cite{Mar:1999}, it is clear that for a fixed number
$N=1,2,3,\ldots$, each of the four supermarket models with $N$ identical
repairable servers is stable if $\widetilde{\rho}<1$. This completes the
proof. \textbf{{\rule{0.08in}{0.08in}}}

The following theorem provides a useful property of the double limit of the
expected fraction vector function $\mathbf{u}^{\left(  N\right)  }\left(
t\right)  =\left(  \mathbf{u}_{W}^{\left(  N\right)  }\left(  t\right)
,\mathbf{V}_{R}^{\left(  N\right)  }\left(  t\right)  \right)  $, which is a
key to establish the systems of nonlinear equations satisfied by the fixed point.

\begin{The}
\label{The:Limit}If $\widetilde{\rho}=\rho\left(  1+\alpha/\beta\right)  <1$,
then for each of the four interrelated repairable supermarket models, there
exists a unique double limit%
\[
\pi=\lim_{\substack{N\rightarrow\infty\\t\rightarrow+\infty}}\mathbf{u}%
^{\left(  N\right)  }\left(  t\right)  .
\]
\end{The}

\textbf{Proof: }This proof is given in Appendix A.
\textbf{{\rule{0.08in}{0.08in}}}

In fact, Theorem \ref{The:Limit} also gives%
\[
\pi=\lim_{N\rightarrow\infty}\lim_{t\rightarrow+\infty}\mathbf{u}^{\left(
N\right)  }\left(  t\right)  =\lim_{t\rightarrow+\infty}\lim_{N\rightarrow
\infty}\mathbf{u}^{\left(  N\right)  }\left(  t\right)  ,
\]
which justifies the interchange of the limit of the expected fraction vector
function $\mathbf{u}^{\left(  N\right)  }\left(  t\right)  $ as $N\rightarrow
\infty$ and $t\rightarrow+\infty$. This is necessary in many practical
applications when using the stationary probabilities to give the effective
approximation for performance of the supermarket models.

Let $\pi=\left(  \pi_{W},\pi_{R}\right)  $, where $\pi_{W}=\left(  \pi
_{W,0},\pi_{W,1},\pi_{W,2},...\right)  $ and $\pi_{R}=\left(  \pi_{R,1,}%
\pi_{R,2,}\pi_{R,3},...\right)  $. The row vector $\pi$ is called a fixed
point of the expected fraction vector function $\mathbf{u}^{\left(  N\right)
}\left(  t\right)  $ if $\pi=\lim_{\substack{N\rightarrow\infty\\t\rightarrow
+\infty}}\mathbf{u}^{\left(  N\right)  }\left(  t\right)  $. Based on Theorem
\ref{The:Limit}, we denote by $\pi_{W,k}=\lim_{\substack{N\rightarrow
\infty\\t\rightarrow+\infty}}u_{W,k}^{\left(  N\right)  }\left(  t\right)  $
for $k\geq0$ and $\pi_{R,l}=\lim_{\substack{N\rightarrow\infty\\t\rightarrow
+\infty}}u_{R,l}^{\left(  N\right)  }\left(  t\right)  $ for $l\geq1$.

It is well-known that if $\pi$ is the fixed point of the expected fraction
vector function $\mathbf{u}^{\left(  N\right)  }\left(  t\right)  $, then%
\[
\lim_{t\rightarrow+\infty}\left[  \frac{\text{d}}{\text{d}t}\mathbf{u}%
^{\left(  N\right)  }\left(  t\right)  \right]  =0,
\]
this gives%
\[
\lim_{t\rightarrow+\infty}\left[  \frac{\text{d}}{\text{d}t}u_{W,k}^{\left(
N\right)  }\left(  t\right)  \right]  =0,k\geq0;\text{ \ }\lim_{t\rightarrow
+\infty}\left[  \frac{\text{d}}{\text{d}t}u_{R,l}^{\left(  N\right)  }\left(
t\right)  \right]  =0,l\geq1.
\]

To set up a system of nonlinear equations, we write%
\[
L_{1}\left(  \pi_{W,0},\pi_{W,1};\pi_{R,1}\right)  =\lim
_{\substack{N\rightarrow\infty\\t\rightarrow+\infty}}L_{1}^{\left(  N\right)
}\left(  u_{W,0},u_{W,1};u_{R,1};t\right)
\]
and for $k\geq2$%
\[
L_{k}\left(  \pi_{W,k-1},\pi_{W,k};\pi_{R,k-1},\pi_{R,k}\right)
=\lim_{\substack{N\rightarrow\infty\\t\rightarrow+\infty}}L_{k}^{\left(
N\right)  }\left(  u_{W,k-1},u_{W,k};u_{R,k-1},u_{R,k};t\right)  ,
\]%
\[
\mathcal{W}_{R,k}^{\left(  N\right)  }\left(  \pi_{W,k-1},\pi_{W,k}%
;\pi_{R,k-1},\pi_{R,k}\right)  =\lim_{\substack{N\rightarrow\infty
\\t\rightarrow+\infty}}\mathcal{W}_{R,k}^{\left(  N\right)  }\left(
u_{W,k-1},u_{W,k};u_{R,k-1},u_{R,k};t\right)  ,
\]%
\[
I_{1}\left(  \pi_{R,1},\pi_{W,2}\right)  =\lim_{\substack{N\rightarrow
\infty\\t\rightarrow+\infty}}I_{1}\left(  u_{R,1}^{\left(  N\right)  }\left(
t\right)  ,u_{W,2}^{\left(  N\right)  };t\right)
\]
and for $k\geq2$%
\[
I_{k}\left(  \pi_{R,k},\pi_{W,k+1}\right)  =\lim_{\substack{N\rightarrow
\infty\\t\rightarrow+\infty}}I_{k}\left(  u_{R,k}^{\left(  N\right)
},u_{W,k+1}^{\left(  N\right)  };t\right)  .
\]
It is easy to check from Theorem \ref{The:Inv} that%
\[
\left(  \pi_{W,0}-\pi_{W,1}\right)  L_{1}\left(  \pi_{W,0},\pi_{W,1};\pi
_{R,1}\right)  =\left(  \pi_{W,0}+\pi_{R,1}\right)  ^{d_{1}}-\left(  \pi
_{W,1}+\pi_{R,1}\right)  ^{d_{1}}%
\]
and for $k\geq2$%
\begin{align*}
&  \left(  \pi_{W,k-1}-\pi_{W,k}\right)  L_{k}\left(  \pi_{W,k-1},\pi
_{W,k};\pi_{R,k-1},\pi_{R,k}\right) \\
&  =\frac{\pi_{W,k-1}-\pi_{W,k}}{\pi_{W,k-1}-\pi_{W,k}+\pi_{R,k-1}-\pi_{R,k}%
}\left[  \left(  \pi_{W,k-1}+\pi_{R,k-1}\right)  ^{d_{1}}-\left(  \pi
_{W,k}+\pi_{R,k}\right)  ^{d_{1}}\right]  ,
\end{align*}%
\begin{align*}
&  \left(  \pi_{R,k-1}-\pi_{R,k}\right)  L_{k}\left(  \pi_{W,k-1},\pi
_{W,k};\pi_{R,k-1},\pi_{R,k}\right) \\
&  =\frac{\pi_{R,k-1}-\pi_{R,k}}{\pi_{W,k-1}-\pi_{W,k}+\pi_{R,k-1}-\pi_{R,k}%
}\left[  \left(  \pi_{W,k-1}+\pi_{R,k-1}\right)  ^{d_{1}}-\left(  \pi
_{W,k}+\pi_{R,k}\right)  ^{d_{1}}\right]  ,
\end{align*}%
\[
\left(  \pi_{R,k-1}-\pi_{R,k}\right)  \mathcal{W}_{R,k}^{\left(  N\right)
}\left(  \pi_{W,k-1},\pi_{W,k};\pi_{R,k-1},\pi_{R,k}\right)  =\left(
\pi_{R,k-1}+\pi_{W,k}\right)  ^{d_{1}}-\left(  \pi_{R,k}+\pi_{W,k}\right)
^{d_{1}},
\]%
\[
I_{1}\left(  \pi_{R,1},\pi_{W,2}\right)  =\left(  1-\pi_{W,2}\right)  ^{d_{2}%
}-\left(  1-\pi_{R,1}-\pi_{W,2}\right)  ^{d_{2}}%
\]
and for $k\geq2$%
\[
I_{k}\left(  \pi_{R,k},\pi_{W,k+1}\right)  =\left(  1-\pi_{W,k+1}\right)
^{d_{2}}-\left(  1-\pi_{R,k}-\pi_{W,k+1}\right)  ^{d_{2}}.
\]

\subsection{Model I ((A.$1$)\textbf{ }and (R.$1$))}

Taking $N\rightarrow\infty$ and $t\rightarrow+\infty$ in both sides of the
mean-field equations (\ref{Equ8}) to (\ref{Equ13}), it is easy to see that the
fixed point satisfies the following system of nonlinear equations%
\begin{equation}
-\alpha\pi_{W,1}+\beta\pi_{R,1}=0, \label{E2}%
\end{equation}%
\begin{equation}
\lambda\left[  \left(  \pi_{W,0}+\pi_{R,1}\right)  ^{d_{1}}-\left(  \pi
_{W,1}+\pi_{R,1}\right)  ^{d_{1}}\right]  -\mu\left(  \pi_{W,1}-\pi
_{W,2}\right)  -\alpha\pi_{W,1}+\beta\pi_{R,1}=0, \label{E3}%
\end{equation}
for $k\geq2$%
\begin{align}
&  \lambda\frac{\pi_{W,k-1}-\pi_{W,k}}{\pi_{W,k-1}-\pi_{W,k}+\pi_{R,k-1}%
-\pi_{R,k}}\left[  \left(  \pi_{W,k-1}+\pi_{R,k-1}\right)  ^{d_{1}}-\left(
\pi_{W,k}+\pi_{R,k}\right)  ^{d_{1}}\right] \nonumber\\
&  -\mu\left(  \pi_{W,k}-\pi_{W,k+1}\right)  -\alpha\pi_{W,k}+\beta\pi
_{R,k}=0, \label{E4}%
\end{align}
and%
\begin{align}
&  \lambda\frac{\pi_{R,k-1}-\pi_{R,k}}{\pi_{W,k-1}-\pi_{W,k}+\pi_{R,k-1}%
-\pi_{R,k}}\left[  \left(  \pi_{W,k-1}+\pi_{R,k-1}\right)  ^{d_{1}}-\left(
\pi_{W,k}+\pi_{R,k}\right)  ^{d_{1}}\right] \nonumber\\
&  +\alpha\pi_{W,k}-\beta\pi_{R,k}=0, \label{E5}%
\end{align}
with the boundary condition%
\begin{equation}
\pi_{W,0}+\pi_{R,1}=1. \label{E6}%
\end{equation}

To solve the system of nonlinear equations (\ref{E2}) to (\ref{E6}), the
following lemma determines the boundary values $\pi_{W,0}$, $\pi_{W,1}$ and
$\pi_{R,1}$, which are a key in our computation of the fixed point later.

\begin{Lem}
\label{Lem:BV}If $\widetilde{\rho}<1$, then
\begin{equation}
\pi_{W,1}=\frac{\lambda}{\mu}=\rho, \label{E7}%
\end{equation}%
\begin{equation}
\pi_{R,1}=\frac{\alpha}{\beta}\rho\label{E7-1}%
\end{equation}
and%
\begin{equation}
\pi_{W,0}=1-\frac{\alpha}{\beta}\rho. \label{E7-2}%
\end{equation}
\end{Lem}

\textbf{Proof}: It follows from (\ref{E2}) and (\ref{E6}) that%
\[
\pi_{R,1}=\frac{\alpha}{\beta}\pi_{W,1}%
\]
and%
\[
\pi_{W,0}=1-\frac{\alpha}{\beta}\pi_{W,1}.
\]
It follows from (\ref{E3}) to (\ref{E5}) that%
\begin{align*}
\pi_{W,1}  &  =\rho\left\{  \left(  \pi_{W,0}+\pi_{R,1}\right)  ^{d_{1}%
}-\left(  \pi_{W,1}+\pi_{R,1}\right)  ^{d_{1}}\right. \\
&  +\left.  \sum_{k=2}^{\infty}\left[  \left(  \pi_{W,k-1}+\pi_{R,k-1}\right)
^{d_{1}}-\left(  \pi_{W,k}+\pi_{R,k}\right)  ^{d_{1}}\right]  \right\} \\
&  =\rho\left(  \pi_{W,0}+\pi_{R,1}\right)  ^{d_{1}}=\rho,
\end{align*}
since $\pi_{W,0}+\pi_{R,1}=1$. This gives%
\[
\pi_{R,1}=\frac{\alpha}{\beta}\rho
\]
and%
\[
\pi_{W,0}=1-\frac{\alpha}{\beta}\rho.
\]
This completes the proof. \textbf{{\rule{0.08in}{0.08in}}}

Let $\xi_{0}=1-\rho\alpha/\beta$, $\xi_{1}=\rho$ and $\delta_{1}=\rho
\alpha/\beta$. Using (\ref{E2}) and (\ref{E6}), we take that $\pi_{W,0}%
=\xi_{0}$, $\pi_{W,1}=\xi_{1}$ and $\pi_{R,1}=\delta_{1}$. Let%
\begin{equation}
\xi_{2}=\xi_{1}-\rho\left(  \xi_{0}-\xi_{1}\right)  L_{1}\left(  \xi_{0}%
,\xi_{1};\delta_{1}\right)  +\frac{\alpha}{\mu}\xi_{1}-\frac{\beta}{\mu}%
\delta_{1} \label{E8}%
\end{equation}
and $\delta_{2}$ the unique solution in $\left(  0,\delta_{1}\right)  $ to the
nonlinear equation%
\begin{equation}
F_{2}\left(  x\right)  =\beta x-\lambda\left(  \delta_{1}-x\right)
L_{2}\left(  \xi_{1},\xi_{2};\delta_{1},x\right)  -\alpha\xi_{2}=0. \label{E9}%
\end{equation}
For $l\geq3$, we set%
\begin{equation}
F_{l}\left(  x\right)  =\beta x-\lambda\left(  \delta_{l-1}-x\right)
L_{l}\left(  \xi_{l-1},\xi_{l};\delta_{l-1},x\right)  -\alpha\xi_{l}.
\label{E10}%
\end{equation}
We assume that for $l\leq k-1$, the $k-1$ pairs $\left(  \xi_{0},\delta
_{1}\right)  ,\left(  \xi_{1},\delta_{2}\right)  ,...,\left(  \xi_{k-2}%
,\delta_{k-1}\right)  $ have been given iteratively, where $\delta_{k-1}$ is
the unique solution in $\left(  0,\delta_{k-2}\right)  $ to the nonlinear
equation $F_{k-1}\left(  x\right)  =0$. For $l=k$, we write%
\begin{equation}
\xi_{k}=\xi_{k-1}-\rho\left(  \xi_{k-2}-\xi_{k-1}\right)  L_{k-1}\left(
\xi_{k-2},\xi_{k-1};\delta_{k-2},\delta_{k-1}\right)  +\frac{\alpha}{\mu}%
\xi_{k-1}-\frac{\beta}{\mu}\delta_{k-1}, \label{E13}%
\end{equation}
and $\delta_{k}$ is the unique solution in $\left(  0,\delta_{k-1}\right)  $
to the nonlinear equation $F_{k}\left(  x\right)  =0$. It is clear that
$0<\xi_{k}<\xi_{k-1}<\cdots<\xi_{1}<\xi_{0}=1-\rho\alpha/\beta$ and
$0<\delta_{k}<\delta_{k-1}<\cdots<\delta_{2}<\delta_{1}=\rho\alpha/\beta.$

The following theorem provides expression for the fixed point by means of the
system of nonlinear equations (\ref{E2}) to (\ref{E6}).

\begin{The}
If $\widetilde{\rho}<1,$ then the fixed point $\pi=\left(  \pi_{W,0},\pi
_{W,1},\pi_{W,2},...;\pi_{R,1},\pi_{R,2},\pi_{R,3},...\right)  $ is given by%
\[
\pi_{W,k}=\xi_{k},k\geq0,
\]
and%
\[
\pi_{R,l}=\delta_{l},l\geq1.
\]
\end{The}

\textbf{Proof: }Lemma \ref{Lem:BV} shows that $\pi_{W,0}=\xi_{0}$, $\pi
_{W,1}=\xi_{1}$ and $\pi_{R,1}=\delta_{1}$.

We assume that for $1\leq l\leq k,\pi_{W,l}=\xi_{l}$ and $\pi_{R,l}=\delta
_{l}$, where $0<\xi_{k}<\xi_{k-1}<\cdots<\xi_{1}<\xi_{0}=1-\rho\alpha/\beta$
and $0<\delta_{k}<\delta_{k-1}<\cdots<\delta_{2}<\delta_{1}=\rho\alpha/\beta.$
Then for $l=k+1$, it follows from Equation (\ref{E3}) that%
\begin{align*}
\pi_{W,k+1}  &  =\pi_{W,k}-\rho\left(  \pi_{W,k-1}-\pi_{W,k}\right)
L_{k}\left(  \pi_{W,k-1},\pi_{W,k};\pi_{R,k-1},\pi_{R,k}\right)
+\frac{\alpha}{\mu}\pi_{W,k}-\frac{\beta}{\mu}\pi_{R,k}\\
&  =\xi_{k}-\rho\left(  \xi_{k-1}-\xi_{k}\right)  L_{k}\left(  \xi_{k-1}%
,\xi_{k};\delta_{k-1},\delta_{k}\right)  +\frac{\alpha}{\mu}\xi_{k}%
-\frac{\beta}{\mu}\delta_{k}=\xi_{k+1.}%
\end{align*}
It follows from Equation (\ref{E5}) that%
\[
\lambda\left(  \delta_{k}-\pi_{R,k+1}\right)  L_{k+1}\left(  \xi_{k},\xi
_{k+1};\delta_{k},\pi_{R,k+1}\right)  +\alpha\xi_{k+1}-\beta\pi_{R,k+1}=0.
\]
Let%
\begin{align*}
F_{k+1}\left(  x\right)   &  =\beta x-\lambda\left(  \delta_{k}-x\right)
L_{k+1}\left(  \xi_{k},\xi_{k+1};\delta_{k},x\right)  -\alpha\xi_{k+1}\\
&  =\beta x-\frac{\lambda\left(  \delta_{k}-x\right)  }{\xi_{k}-\xi
_{k+1}+\delta_{k}-x}\left[  \left(  \xi_{k}+\delta_{k}\right)  ^{d_{1}%
}-\left(  \xi_{k+1}+x\right)  ^{d_{1}}\right]  -\alpha\xi_{k+1}.
\end{align*}
Then
\[
F_{k+1}\left(  0\right)  =-\frac{\lambda\delta_{k}}{\xi_{k}-\xi_{k+1}%
+\delta_{k}}\left[  \left(  \xi_{k}+\delta_{k}\right)  ^{d_{1}}-\xi
_{k+1}^{d_{1}}\right]  -\alpha\xi_{k}<0,
\]%
\[
F_{k+1}\left(  \delta_{k}\right)  =\beta\delta_{k}-\alpha\xi_{k+1}>\beta
\delta_{k}-\alpha\xi_{k}=\lambda\left(  \delta_{k-1}-\delta_{k}\right)
L_{k}\left(  \xi_{k-1},\xi_{k};\delta_{k-1},\delta_{k}\right)  >0
\]
by means of (\ref{E5}), and%
\begin{align*}
\frac{\text{d}}{\text{d}x}F_{k+1}\left(  x\right)  =  &  \beta-\frac{\text{d}%
}{\text{d}x}\left[  \frac{\lambda\left(  \delta_{k}-x\right)  \left(  \xi
_{k}+\delta_{k}\right)  ^{d_{1}}}{\xi_{k}-\xi_{k+1}+\delta_{k}-x}\right]
+\frac{\text{d}}{\text{d}x}\left[  \frac{\lambda\left(  \delta_{k}-x\right)
\left(  \xi_{k+1}+x\right)  ^{d_{1}}}{\xi_{k}-\xi_{k+1}+\delta_{k}-x}\right]
\\
=  &  \beta+\lambda\frac{\left(  \xi_{k}+\delta_{k}\right)  ^{d_{1}}\left(
\xi_{k}-\xi_{k+1}\right)  }{\left(  \xi_{k}-\xi_{k+1}+\delta_{k}-x\right)
^{2}}-\lambda\frac{\left(  \xi_{k+1}+x\right)  ^{d_{1}}\left(  \xi_{k}%
-\xi_{k+1}\right)  }{\left(  \xi_{k}-\xi_{k+1}+\delta_{k}-x\right)  ^{2}}\\
&  +\lambda\frac{d_{1}\left(  \xi_{k+1}+x\right)  ^{d_{1}-1}\left(  \delta
_{k}-x\right)  \left(  \xi_{k}-\xi_{k+1}+\delta_{k}-x\right)  }{\left(
\xi_{k}-\xi_{k+1}+\delta_{k}-x\right)  ^{2}}\\
\geq &  \beta+\lambda\frac{d_{1}\left(  \xi_{k+1}+x\right)  ^{d_{1}-1}\left(
\delta_{k}-x\right)  \left(  \xi_{k}-\xi_{k+1}+\delta_{k}-x\right)  }{\left(
\xi_{k}-\xi_{k+1}+\delta_{k}-x\right)  ^{2}}>0
\end{align*}
by means of $\xi_{k}+\delta_{k}\geq\xi_{k+1}+x$. Note that $F_{k+1}\left(
x\right)  $ is a continuous function for $x\in\left(  0,\delta_{k}\right)  $,
there exists a unique positive solution $\delta_{k+1}$ in $\left(
0,\delta_{k}\right)  $ to the nonlinear equation $F_{k+1}\left(  x\right)
=0$. Hence, $\pi_{R,k+1}=\delta_{k+1}.$

By induction, this completes the proof. \textbf{{\rule{0.08in}{0.08in}}}

Note that for the other three models with more complex complex nonlinear
structures, we provide some discussion on the boundary conditions: $\pi
_{W,0},\pi_{W,1}$ and $\pi_{R,1}$.

\subsection{Model II ((A.$1$)\textbf{ }and (R.$2$))}

Taking $N\rightarrow\infty$ and $t\rightarrow+\infty$ in both sides of the
mean-field equations (\ref{MII-1}) to (\ref{MII-5}), it is easy to see that
the fixed point satisfies the following system of nonlinear equations%
\begin{equation}
-\alpha\pi_{W,1}+\beta\left[  \left(  1-\pi_{W,2}\right)  ^{d_{2}}-\left(
1-\pi_{R,1}-\pi_{W,2}\right)  ^{d_{2}}\right]  =0, \label{MII-F1}%
\end{equation}%
\begin{align}
&  \lambda\left[  \left(  \pi_{W,0}+\pi_{R,1}\right)  ^{d_{1}}-\left(
\pi_{W,1}+\pi_{R,1}\right)  ^{d_{1}}\right]  -\mu\left(  \pi_{W,1}-\pi
_{W,2}\right)  -\alpha\pi_{W,1}\nonumber\\
&  +\beta\left[  \left(  1-\pi_{W,2}\right)  ^{d_{2}}-\left(  1-\pi_{R,1}%
-\pi_{W,2}\right)  ^{d_{2}}\right]  =0, \label{MII-F2}%
\end{align}
for $k\geq2$%
\begin{align}
&  \lambda\frac{\pi_{W,k-1}-\pi_{W,k}}{\pi_{W,k-1}-\pi_{W,k}+\pi_{R,k-1}%
-\pi_{R,k}}\left[  \left(  \pi_{W,k-1}+\pi_{R,k-1}\right)  ^{d_{1}}-\left(
\pi_{W,k}+\pi_{R,k}\right)  ^{d_{1}}\right] \nonumber\\
&  -\mu\left(  \pi_{W,k}-\pi_{W,k+1}\right)  -\alpha\pi_{W,k}+\beta\left[
\left(  1-\pi_{W,k+1}\right)  ^{d_{2}}-\left(  1-\pi_{R,k}-\pi_{W,k+1}\right)
^{d_{2}}\right]  =0, \label{MII-F3}%
\end{align}
and%
\begin{align}
&  \lambda\frac{\pi_{R,k-1}-\pi_{R,k}}{\pi_{W,k-1}-\pi_{W,k}+\pi_{R,k-1}%
-\pi_{R,k}}\left[  \left(  \pi_{W,k-1}+\pi_{R,k-1}\right)  ^{d_{1}}-\left(
\pi_{W,k}+\pi_{R,k}\right)  ^{d_{1}}\right] \nonumber\\
&  +\alpha\pi_{W,k}-\beta\left[  \left(  1-\pi_{W,k+1}\right)  ^{d_{2}%
}-\left(  1-\pi_{R,k}-\pi_{W,k+1}\right)  ^{d_{2}}\right]  =0, \label{MII-F4}%
\end{align}
with the boundary condition%
\begin{equation}
\pi_{W,0}+\pi_{R,1}=1. \label{MII-F5}%
\end{equation}

It follows from (\ref{MII-F1}) and (\ref{MII-F2}) that%
\[
\lambda\left[  \left(  \pi_{W,0}+\pi_{R,1}\right)  ^{d_{1}}-\left(  \pi
_{W,1}+\pi_{R,1}\right)  ^{d_{1}}\right]  -\mu\left(  \pi_{W,1}-\pi
_{W,2}\right)  =0,
\]
and from (\ref{MII-F3}) and (\ref{MII-F4}) that for $k\geq2$%
\[
\lambda\left[  \left(  \pi_{W,k-1}+\pi_{R,k-1}\right)  ^{d_{1}}-\left(
\pi_{W,k}+\pi_{R,k}\right)  ^{d_{1}}\right]  -\mu\left(  \pi_{W,k}-\pi
_{W,k+1}\right)  =0,
\]
which, together with (\ref{MII-F5}), follows%
\begin{equation}
\pi_{W,1}=\rho\left(  \pi_{W,0}+\pi_{R,1}\right)  ^{d_{1}}=\rho.
\label{MII-F6}%
\end{equation}
It follows from (\ref{MII-F1}), (\ref{MII-F2}) and (\ref{MII-F6})\ that%
\begin{equation}
\pi_{W,2}=\rho\left(  \rho+\pi_{R,1}\right)  ^{d_{1}}. \label{MII-F7}%
\end{equation}
From (\ref{MII-F6}), (\ref{MII-F7}) and (\ref{MII-F2}), we find that
$\pi_{R,1}$ is the minimal nonnegative solution to the following nonlinear
equation
\[
\left[  1-\rho\left(  \rho+\pi_{R,1}\right)  ^{d_{1}}\right]  ^{d_{2}}-\left[
1-\pi_{R,1}-\rho\left(  \rho+\pi_{R,1}\right)  ^{d_{1}}\right]  ^{d_{2}}%
=\rho\frac{\alpha}{\beta}.
\]
Also, $\pi_{W,0}=1-\pi_{R,1}$ is given.

\subsection{Model III ((A.$2$)\textbf{ }and (R.$1$))}

Taking $N\rightarrow\infty$ and $t\rightarrow+\infty$ in both sides of the
mean-field equations (\ref{MIII-1}) to (\ref{MIII-5}), we obtain that the
fixed point satisfies the following system of nonlinear equations%
\begin{equation}
-\alpha\pi_{W,1}+\beta\pi_{R,1}=0, \label{MIII-F1}%
\end{equation}%
\begin{equation}
\lambda\left[  \left(  \pi_{W,0}+\pi_{R,1}\right)  ^{d_{1}}-\left(  \pi
_{W,1}+\pi_{R,1}\right)  ^{d_{1}}\right]  -\mu\left(  \pi_{W,1}-\pi
_{W,2}\right)  -\alpha\pi_{W,1}+\beta\pi_{R,1}=0, \label{MIII-F2}%
\end{equation}
for $k\geq2$%
\begin{align}
&  \lambda\frac{\pi_{W,k-1}-\pi_{W,k}}{\pi_{W,k-1}-\pi_{W,k}+\pi_{R,k-1}%
-\pi_{R,k}}\left[  \left(  \pi_{W,k-1}+\pi_{R,k-1}\right)  ^{d_{1}}-\left(
\pi_{W,k}+\pi_{R,k}\right)  ^{d_{1}}\right] \nonumber\\
&  -\mu\left(  \pi_{W,k}-\pi_{W,k+1}\right)  -\alpha\pi_{W,k}+\beta\pi
_{R,k}=0, \label{MIII-F3}%
\end{align}
and%
\begin{equation}
\lambda\left[  \left(  \pi_{R,k-1}+\pi_{W,k}\right)  ^{d_{1}}-\left(
\pi_{R,k}+\pi_{W,k}\right)  ^{d_{1}}\right]  +\alpha\pi_{W,k}-\beta\pi
_{R,k}=0, \label{MIII-F4}%
\end{equation}
with the boundary condition%
\begin{equation}
\pi_{W,0}+\pi_{R,1}=1. \label{MIII-F5}%
\end{equation}

Now, we discuss the boundary conditions of the fixed point. It follows from
(\ref{MIII-F1}) and (\ref{MIII-F5}) that%
\[
\pi_{R,1}=\frac{\alpha}{\beta}\pi_{W,1}%
\]
and%
\[
\pi_{W,0}=1-\frac{\alpha}{\beta}\pi_{W,1}.
\]
It follows from (\ref{MIII-F2}) to (\ref{MIII-F4}) that%
\begin{align*}
\frac{1}{\rho}\pi_{W,1}  &  =1+\sum_{k=2}^{\infty}\left[  \left(  \pi
_{R,k-1}+\pi_{W,k}\right)  ^{d_{1}}-\left(  \pi_{R,k}+\pi_{W,k}\right)
^{d_{1}}\right] \\
&  -\sum_{k=2}^{\infty}\frac{\pi_{R,k-1}-\pi_{R,k}}{\pi_{W,k-1}-\pi_{W,k}%
+\pi_{R,k-1}-\pi_{R,k}}\left[  \left(  \pi_{R,k-1}+\pi_{W,k-1}\right)
^{d_{1}}-\left(  \pi_{R,k}+\pi_{W,k}\right)  ^{d_{1}}\right] \\
&  =1+\sum_{k=2}^{\infty}\left[  \left(  \pi_{W,k-1}+\pi_{R,k}\right)
^{d_{1}}\right. \\
&  \left.  -\frac{\left(  \pi_{W,k-1}-\pi_{W,k}\right)  \left(  \pi_{W,k}%
+\pi_{R,k}\right)  ^{d_{1}}+\left(  \pi_{R,k-1}-\pi_{R,k}\right)  \left(
\pi_{W,k-1}+\pi_{R,k-1}\right)  ^{d_{1}}}{\pi_{W,k-1}-\pi_{W,k}+\pi
_{R,k-1}-\pi_{R,k}}\right]  .
\end{align*}
Let%
\begin{align*}
\Delta\left(  d_{1}\right)  =  &  \sum_{k=2}^{\infty}\frac{\left(  \pi
_{W,k-1}-\pi_{W,k}\right)  \left(  \pi_{W,k}+\pi_{R,k}\right)  ^{d_{1}%
}+\left(  \pi_{R,k-1}-\pi_{R,k}\right)  \left(  \pi_{W,k-1}+\pi_{R,k-1}%
\right)  ^{d_{1}}}{\pi_{W,k-1}-\pi_{W,k}+\pi_{R,k-1}-\pi_{R,k}}\\
&  -\sum_{k=2}^{\infty}\left(  \pi_{W,k-1}+\pi_{R,k}\right)  ^{d_{1}}.
\end{align*}
This gives%
\[
\pi_{W,1}=\rho\left[  1-\Delta\left(  d_{1}\right)  \right]  .
\]
It is easy to check that $\Delta\left(  d_{1}\right)  \in\left(  0,1\right)
$, and $\Delta\left(  d_{1}\right)  $ is increasing in $d_{1}\geq1$. Therefore
$\pi_{W,1}<\rho$ if $d_{1}\geq2$, and $\pi_{W,1}=\rho$ if $d_{1}=1$. At the
same time, $\pi_{W,1}$ is decreasing in $d_{1}\geq1$.

\subsection{Model IV ((A.$2$)\textbf{ }and (R.$2$))}

Taking $N\rightarrow\infty$ and $t\rightarrow+\infty$ in both sides of the
mean-field equations (\ref{MIV-1}) to (\ref{MIV-5}), it is easy to see that
the fixed point satisfies the following system of nonlinear equations%
\begin{equation}
-\alpha\pi_{W,1}+\beta\left[  \left(  1-\pi_{W,2}\right)  ^{d_{2}}-\left(
1-\pi_{R,1}-\pi_{W,2}\right)  ^{d_{2}}\right]  =0, \label{MIV-F1}%
\end{equation}%
\begin{align}
&  \lambda\left[  \left(  \pi_{W,0}+\pi_{R,1}\right)  ^{d_{1}}-\left(
\pi_{W,1}+\pi_{R,1}\right)  ^{d_{1}}\right]  -\mu\left(  \pi_{W,1}-\pi
_{W,2}\right)  -\alpha\pi_{W,1}\nonumber\\
&  +\beta\left[  \left(  1-\pi_{W,2}\right)  ^{d_{2}}-\left(  1-\pi_{R,1}%
-\pi_{W,2}\right)  ^{d_{2}}\right]  =0, \label{MIV-F2}%
\end{align}
for $k\geq2$%
\begin{align}
&  \lambda\frac{\pi_{W,k-1}-\pi_{W,k}}{\pi_{W,k-1}-\pi_{W,k}+\pi_{R,k-1}%
-\pi_{R,k}}\left[  \left(  \pi_{W,k-1}+\pi_{R,k-1}\right)  ^{d_{1}}-\left(
\pi_{W,k}+\pi_{R,k}\right)  ^{d_{1}}\right] \nonumber\\
&  -\mu\left(  \pi_{W,k}-\pi_{W,k+1}\right)  -\alpha\pi_{W,k}+\beta\left[
\left(  1-\pi_{W,k+1}\right)  ^{d_{2}}-\left(  1-\pi_{R,k}-\pi_{W,k+1}\right)
^{d_{2}}\right]  =0, \label{MIV-F3}%
\end{align}
and%
\begin{align}
&  \lambda\left[  \left(  \pi_{R,k-1}+\pi_{W,k}\right)  ^{d_{1}}-\left(
\pi_{R,k}+\pi_{W,k}\right)  ^{d_{1}}\right] \nonumber\\
&  +\alpha\pi_{W,k}-\beta\left[  \left(  1-\pi_{W,k+1}\right)  ^{d_{2}%
}-\left(  1-\pi_{R,k}-\pi_{W,k+1}\right)  ^{d_{2}}\right]  =0, \label{MIV-F4}%
\end{align}
with the boundary condition%
\begin{equation}
\pi_{W,0}+\pi_{R,1}=1. \label{MIV-F5}%
\end{equation}

From the similar analysis to the boundary conditions in Model III, we obtain%
\[
\pi_{W,1}=\rho\left[  1-\Delta\left(  d_{1}\right)  \right]  ,
\]
and the similar analysis to that in Model II leads to%
\[
\pi_{W,2}=-\rho\Delta\left(  d_{1}\right)  +\rho\left\{  \rho\left[
1-\Delta\left(  d_{1}\right)  \right]  +\pi_{R,1}\right\}  ^{d_{1}},
\]
and $\pi_{R,1}$ is the minimal nonnegative solution to the following nonlinear
equation%
\[
\left(  1-\pi_{W,2}\right)  ^{d_{2}}-\left(  1-\pi_{R,1}-\pi_{W,2}\right)
^{d_{2}}=\pi_{W,1}\frac{\alpha}{\beta}.
\]
Also, we get that $\pi_{W,0}=1-\pi_{R,1}$.

\section{Performance Analysis and Numerical Observations}

In this section, we first provide useful performance measures of the four
interrelated supermarket models with repairable servers. Then we use some
numerical examples to make valuable observations on model improvement by means
of performance numerical comparison.

\subsection{Performance measures}

\textbf{(a) The mean of stationary queueing length}

Let $\mathcal{Q}$ be the stationary queue length of any server in each of the
four supermarket models. Then%
\begin{align*}
E\left[  \mathcal{Q}\right]   &  =\sum_{k=1}^{\infty}P\left\{  \mathcal{Q}\geq
k\right\} \\
&  =\sum_{k=1}^{\infty}\left\{  P\left\{  \mathcal{Q}\geq k,\text{the server
is working}\right\}  +P\left\{  \mathcal{Q}\geq k,\text{the server is
repairing}\right\}  \right\} \\
&  =\sum_{k=1}^{\infty}\left(  \pi_{W,k}+\pi_{R,k}\right)  .
\end{align*}

\textbf{(b) The variance of stationary queueing length}

It is easy to check that
\[
Var\left[  \mathcal{Q}\right]  =\sum_{k=1}^{\infty}\left(  2k-1\right)
\left(  \pi_{W,k}+\pi_{R,k}\right)  -\left[  \sum_{k=1}^{\infty}\left(
\pi_{W,k}+\pi_{R,k}\right)  \right]  ^{2},
\]
since%
\begin{align*}
E\left[  \mathcal{Q}^{2}\right]   &  =\sum_{k=1}^{\infty}k^{2}\left[
P\left\{  \mathcal{Q}\geq k\right\}  -P\left\{  \mathcal{Q}\geq k+1\right\}
\right] \\
&  =\sum_{k=1}^{\infty}\left(  2k-1\right)  P\left\{  \mathcal{Q}\geq
k\right\}  .
\end{align*}

\textbf{(c) The steady-state availability and failure frequency}

Let $A$ and $W_{f}$ be the steady-state availability and failure frequency in
any repairable server, respectively. Then%
\begin{equation}
A=\sum\limits_{k=0}^{\infty}\left(  \pi_{W,k}-\pi_{W,k+1}\right)  =\pi_{W,0}
\label{equation4}%
\end{equation}
and%
\begin{equation}
W_{f}=\alpha\sum\limits_{k=1}^{\infty}\left(  \pi_{W,k}-\pi_{W,k+1}\right)
=\alpha\pi_{W,1}. \label{equation5}%
\end{equation}

\textbf{(d) The steady-state mean-field flow balancing}

Since the mean-field theory plays an important role in the study of
supermarket models, a flow balancing in the supermarket models is
called a mean-field flow balancing. In every supermarket model, the
steady-state
mean-field throughput is given by%
\[
\text{MF-TH}=\mu\pi_{W,1}.
\]
Thus the the steady-state mean-field input-output difference is defined as%
\[
F\left(  d_{1},d_{2}\right)  =\lambda-\mu\pi_{W,1}.
\]
Clearly, this supermarket model can have a steady-state mean-field flow
balancing if $F\left(  d_{1},d_{2}\right)  =0$.

\subsection{Numerical observations}

Now, we use some numerical examples to show how the major performance measures
depend on some crucial parameters of the systems. These numerical examples are
organized in three groups for different purposes: (1) examining the mean
$E\left[  \mathcal{Q}\right]  $ and the variance $Var\left[  \mathcal{Q}%
\right]  $; (2) observing the availability $A$ and the failure frequency
$W_{f}$; and (3) discussing the mean-field flow balancing $F\left(
d_{1},d_{2}\right)  $. At the same time, it is worth noting that in the
numerical examples, (A.$1$) and (A.$2$) represent routing customers and
(R.$1$) and (R.$2$) represent organizing of repair resource.

The following seven numerical examples are based on a set of system parameters
of $\mu=9$, $\alpha=2$, $\beta=5$ and $\lambda\in(0,6)$. It is easy to check
that $\widetilde{\rho}<42/45<1$.

\textbf{Example 1: }Show $E\left[  \mathcal{Q}\right]  $ in Models III and IV
for a comparison of deployment of repair resource. In this example with
(A.$2$), we consider Models III and IV and observe how choice numbers $d_{1}$
and $d_{2}$ affect $E\left[  \mathcal{Q}\right]  $, the stationary queue
length. Figure 3 shows how $E\left[  \mathcal{Q}\right]  $ changes with
$d_{1},d_{2}=1,2,3$ when the arrival rate $\lambda\in(0,6)$. It is observed
that while $E\left[  \mathcal{Q}\right]  $ increases with $\lambda$, it
decreases with either $d_{1}$ or $d_{2}$. Also, $d_{1}$ is more effective than
$d_{2}$ in terms of reducing $E\left[  \mathcal{Q}\right]  $.

\begin{figure}[ptbh]
\centering               \includegraphics[width=10cm]{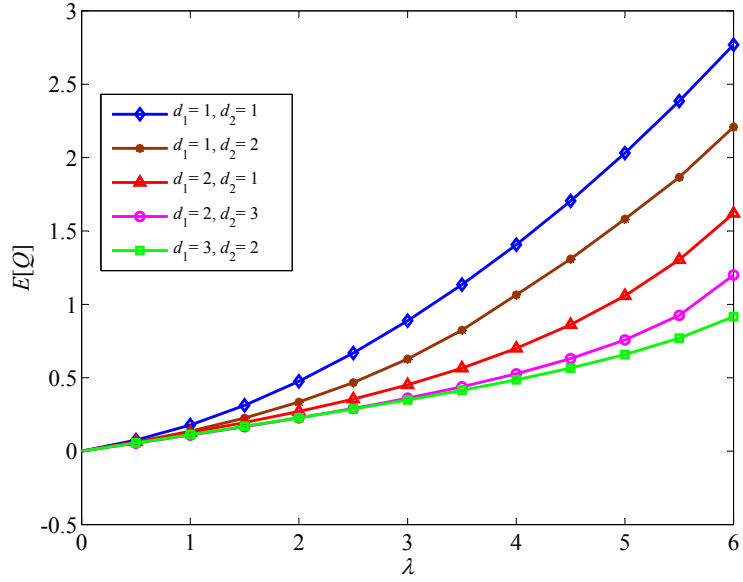}
\caption{$E\left[  \mathcal{Q}\right]  $ for a comparison of repair
organizations}%
\label{figure: fig-3}%
\end{figure}

\textbf{Example 2: } Focus on $E\left[  \mathcal{Q}\right]  $ to compare
(A.$1$) with (A.$2$)

In this example, we demonstrate how (A.$2$) improves the performance under
(A.$1$) in terms of $E\left[  \mathcal{Q}\right]  $. Figure 4 shows the
$E\left[  \mathcal{Q}\right]  $ as a function of the arrival rate $\lambda
\in(0,6)$ with $d_{1},d_{2}=2,3$. It is observed that (A.$2$) can effectively
reduce $E\left[  \mathcal{Q}\right]  $ compared with (A.$1$). This implies
that using more system information can improve the system performance.

\begin{figure}[ptbh]
\centering                \includegraphics[width=7cm]{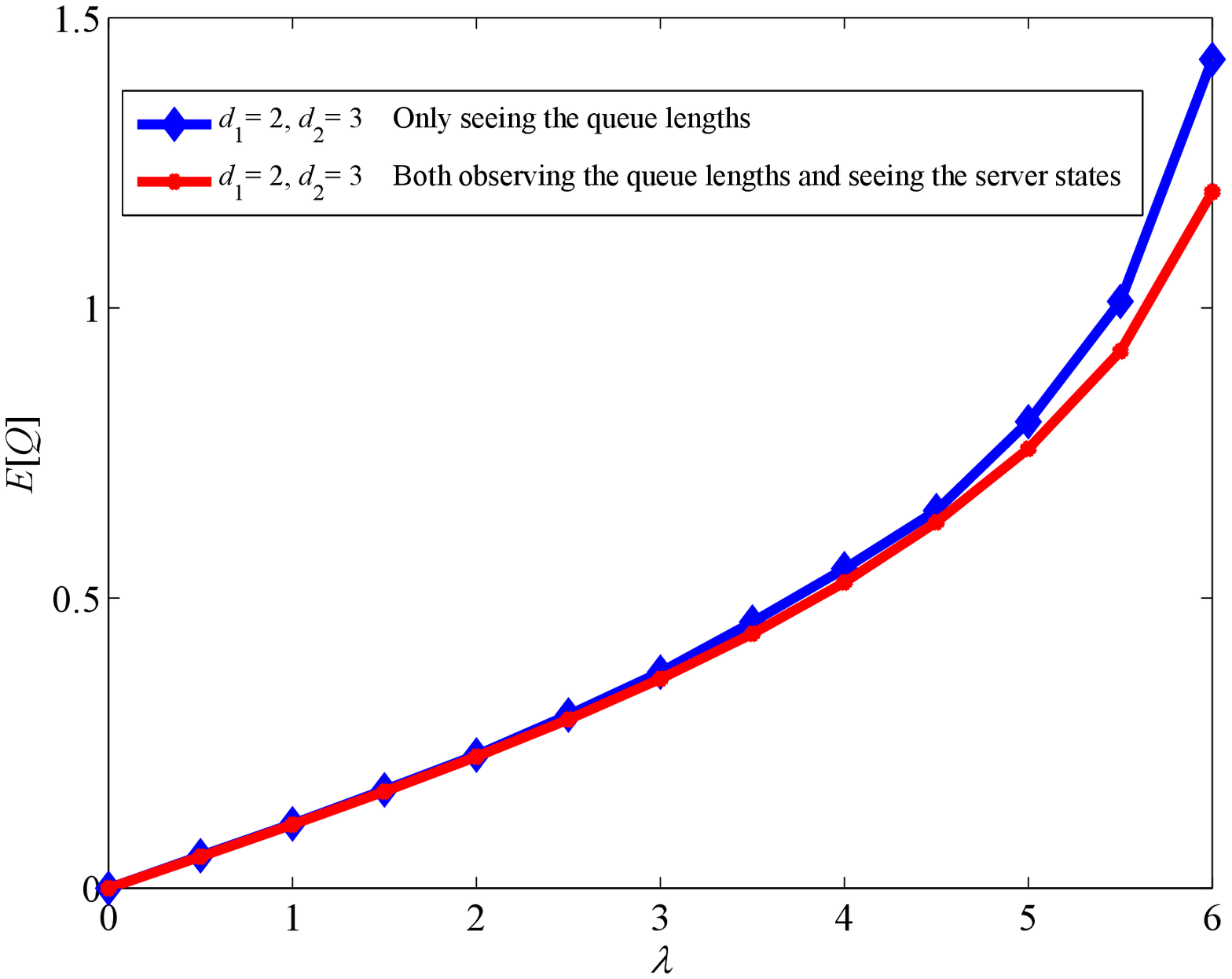}
\includegraphics[width=7cm]{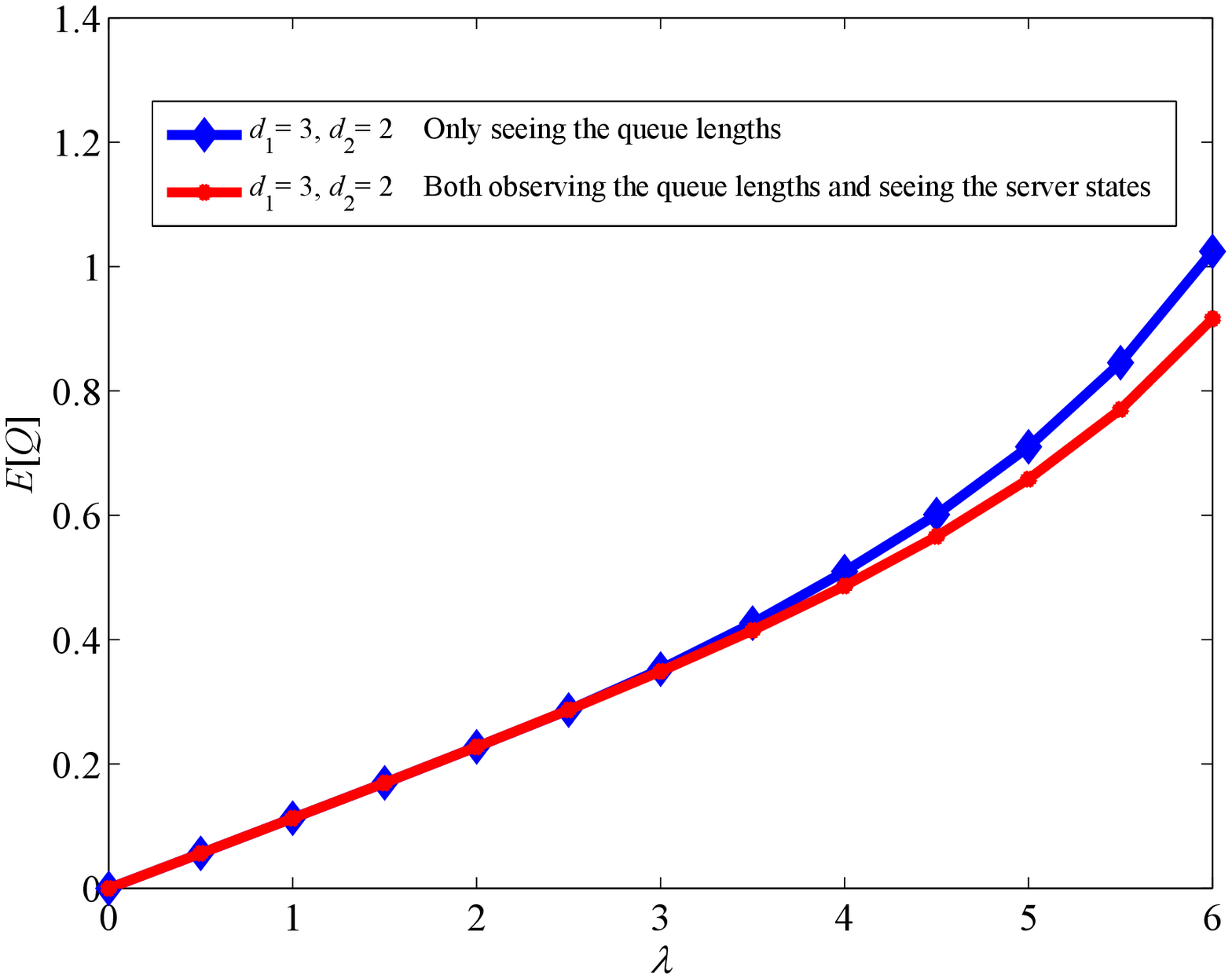}  \caption{$E\left[  \mathcal{Q}%
\right]  $ for comparing (A.$1$) with (A.$2$)}%
\label{figure: fig-4}%
\end{figure}

\textbf{Example 3: } Show $Var\left[  \mathcal{Q}\right]  $ in Models III and
IV for comparing repair resource deployment.

In this example with (A.$2$), we focus on how $d_{1}$ and $d_{2}$ affect
$Var\left[  \mathcal{Q}\right]  $, the variance of stationary queue length.
Figure 5 illustrates how $Var\left[  \mathcal{Q}\right]  $ changes with the
arrival rate $\lambda\in(0,6)$ with $d_{1},d_{2}=1,2,3$. We observe that the
mean queue length decreases with either $d_{1}$ or $d_{2}$. Also, $d_{1}$ is
more effective than $d_{2}$ in terms of reducing $Var\left[  \mathcal{Q}%
\right]  $.

\begin{figure}[ptbh]
\centering               \includegraphics[width=10cm]{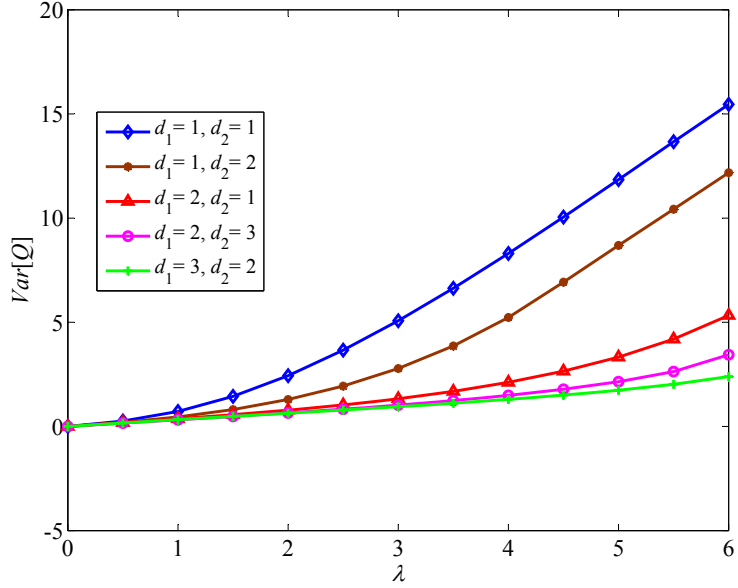}
\caption{$Var\left[  \mathcal{Q}\right]  $ for a comparison of repair
organizations}%
\label{figure: fig-5}%
\end{figure}

\textbf{Example 4: } Observe $Var\left[  \mathcal{Q}\right]  $ under (A.$1$)
or (A.$2$)

In this example, Figure 6 shows how $Var\left[  \mathcal{Q}\right]  $ changes
on the arrival rate $\lambda\in(0,6)$ with $d_{1},d_{2}=2,3$. It is revealed
that (A.$2$) can effectively reduce $Var\left[  \mathcal{Q}\right]  $ under
(A.$1$).

\begin{figure}[ptbh]
\centering                \includegraphics[width=7cm]{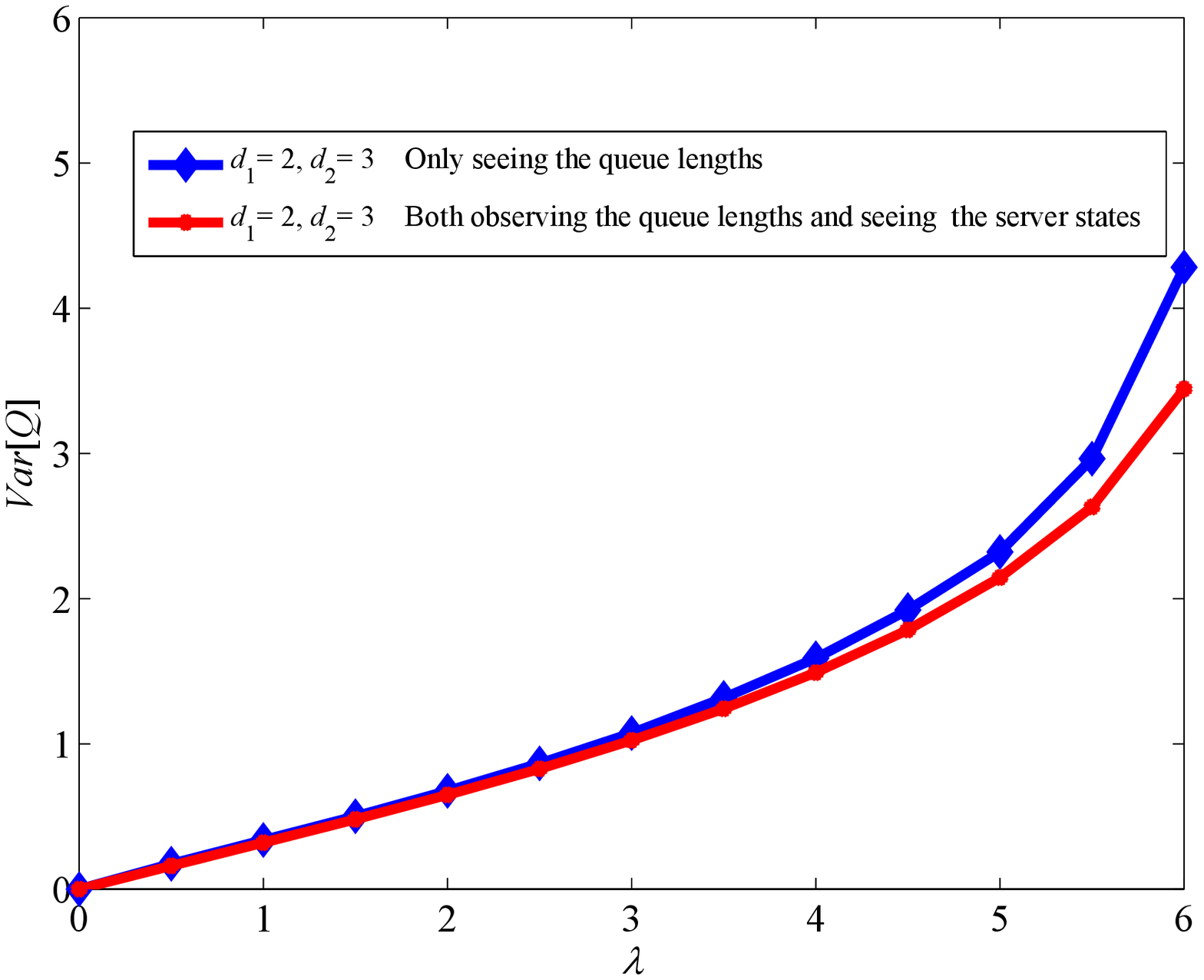}
\includegraphics[width=7cm]{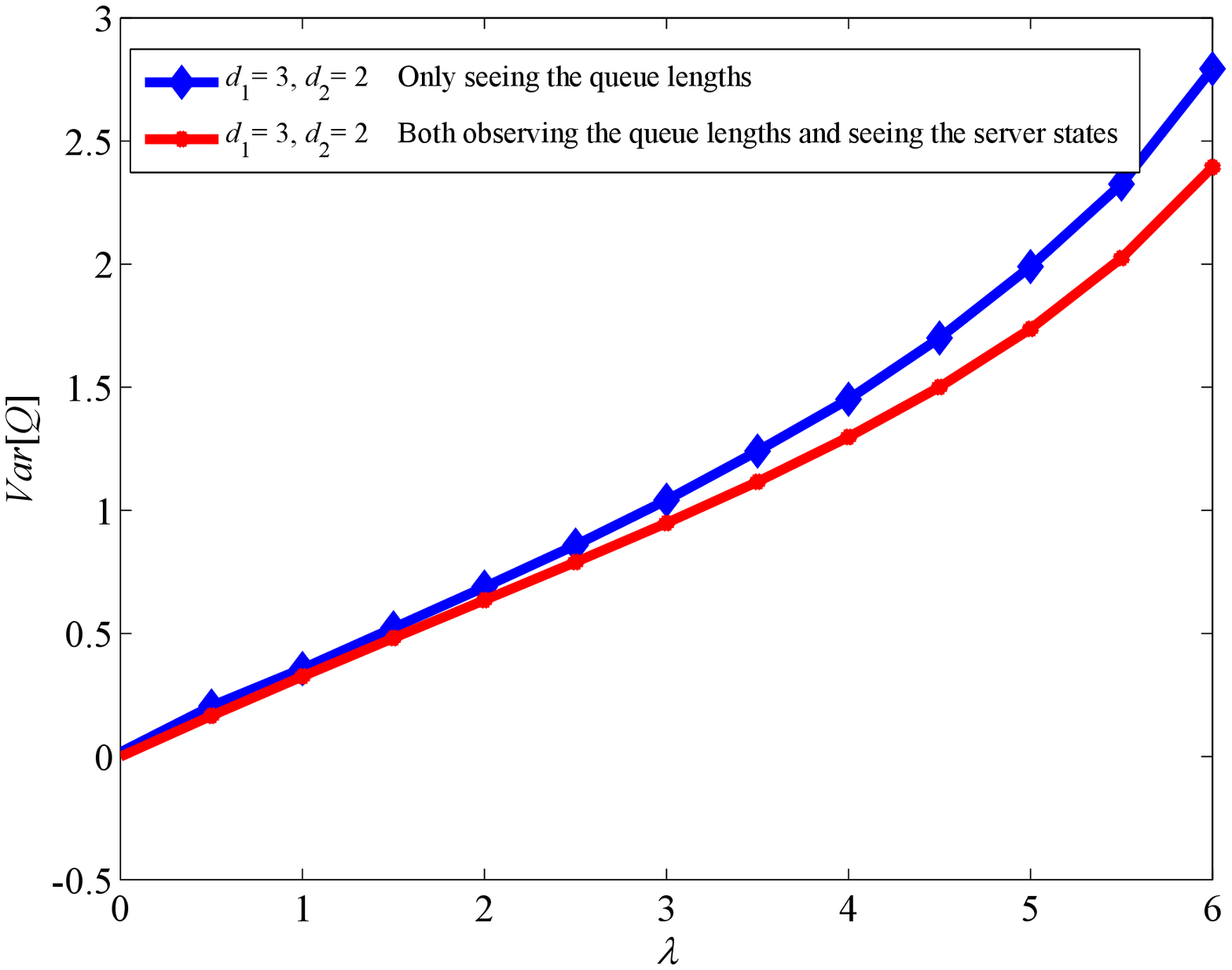}  \caption{$Var\left[  \mathcal{Q}%
\right]  $ for comparing (A.$1$) with (A.$2$)}%
\label{figure: fig-6}%
\end{figure}

\textbf{Example 5:} Examine the steady-state availability $A$ in Models III
and IV

In this example with (A.$2$), Figure 7 shows that while the steady-state
availability $A$ decreases with $\lambda$, it increases with either $d_{1}$ or
$d_{2}$. Thus, $d_{1}$ and $d_{2}$ can help increase the steady-state availability.

\begin{figure}[ptbh]
\centering              \includegraphics[width=9cm]{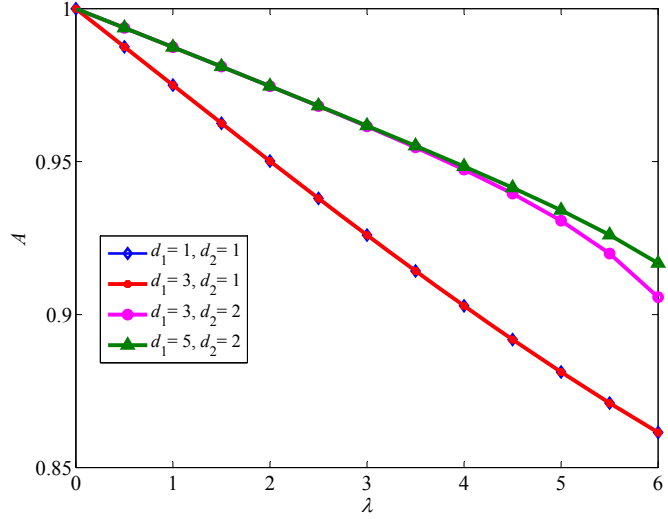}  \caption{$A$
in Models III and IV}%
\label{figure: fig-7}%
\end{figure}

\textbf{Example 6:} Investigate the steady-state failure frequency $W_{f}$ in
Models III and IV

In this example with (A.$2$), Figure 8 shows that $W_{f}$ increases with both
$\lambda$ and $d_{1}$ or $d_{2}$.

\begin{figure}[ptbh]
\centering              \includegraphics[width=9cm]{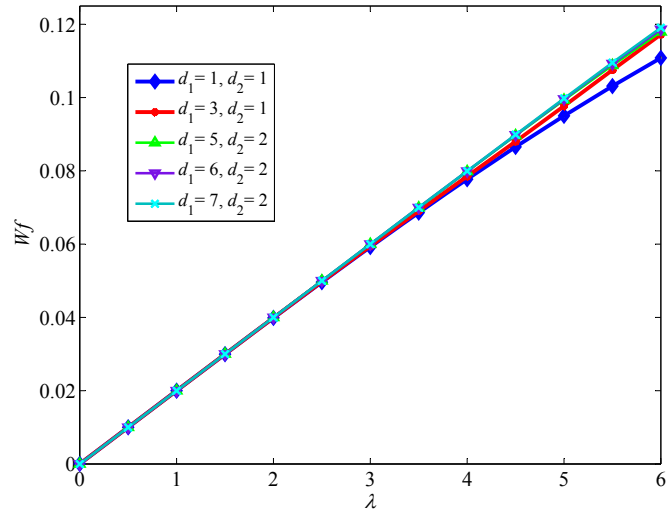}
\caption{$W_{f}$ in Models III and IV}%
\label{figure: fig-8}%
\end{figure}

Finally, we provide a numerical example to show the steady-state flow
balancing in the study of supermarket models.

\textbf{Example 7:} Observe the steady-state mean-field flow balancing

In this example with (A.$2$), we show how the steady-state mean-field
input-output difference $F\left(  d_{1},d_{2}\right)  $ depends on the arrival
rate $\lambda\in(0,5)$ with $d_{1}=1,5,6$ and $d_{2}=1,2,3$.

Figure 9 indicates that if $d_{1}=5,6$ and $d_{2}=2,3$, the steady-state
mean-field input-output difference $F\left(  d_{1},d_{2}\right)  >0$, and it
increases with $\lambda$. However, $F\left(  1,1\right)  =0$, which implies
that the repairable supermarket model has the steady-state mean-field flow
balancing for $d_{1},d_{2}=1$.

\begin{figure}[ptbh]
\centering               \includegraphics[width=11cm]{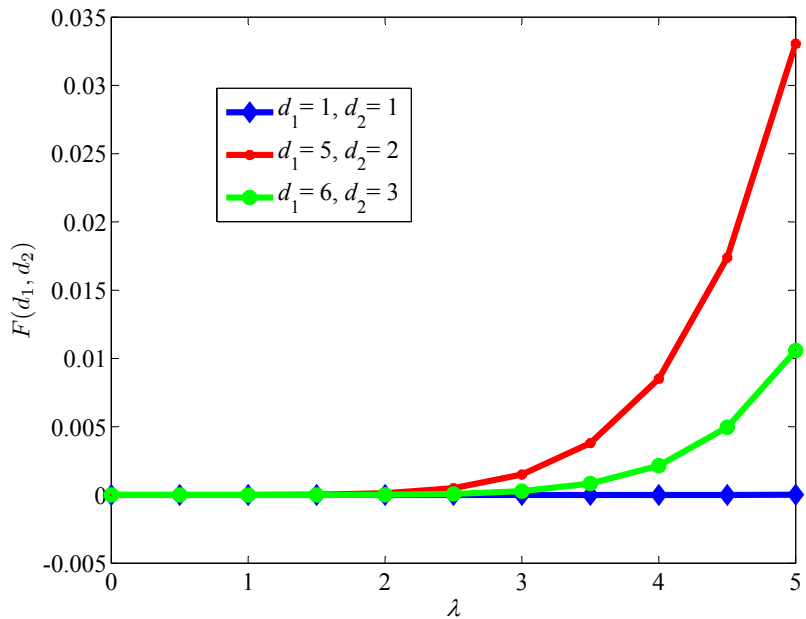}
\caption{$F\left(  d_{1},d_{2}\right)  $ in Models III and IV}%
\label{figure: fig-9}%
\end{figure}

From the numerical analysis above, we may conclude that the system information
(i.e., server in working or repair condition and queue length) for the
arriving customer and the deployment of the repair resource can effectively
improve the system performances of the supermarket models.

\section{Concluding Remarks}

In this paper, we apply the mean-field theory to studying effects of
a double dynamic routine selection scheme (for the arrival
dispatched schemes and for the groups of repairmen) on performance
of the four interrelated supermarket models with repairable servers.
We first provide a probability method of setting up the
infinite-dimensional systems of mean-field equations. Then we prove
asymptotic independence of the supermarket models with repairable
servers. Based on this, we discuss the fixed points which are
computed by means of the systems of nonlinear equations. Finally, we
provide useful performance measures of the supermarket models, and
use some numerical examples to make valuable observations on model
improvement via using system information and deploying repair
resource. Our results reveal effects of utilizing system information
for customer's joining decisions as well as reorganization of repair
resource\ on performance of the supermarket models. Along with this
line, there are a number of interesting directions for future
research, for example:

\begin{itemize}
\item analyzing non-Poisson inputs, such as, Markovian arrival processes
(MAPs), and renewal processes;

\item studying non-exponential service time distributions, for example,
general distributions, matrix-exponential distributions and heavy-tailed distributions;

\item discussing the bulk arrival processes, and the bulk service processes; and

\item developing effective algorithms for computing the fixed points in the
study of complex supermarket models.
\end{itemize}

\section*{Acknowledgements}

The first two authors were supported by the National Natural Science
Foundation of China under grant No. 71471160, No. 71671158 and No. 71471114,
and the Fostering Plan of Innovation Team and Leading Talent in Hebei
Universities under grant No. LJRC027.

\section*{Appendix A: The Proof of Theorem \ref{The:Limit}}

In this appendix, for the four interrelated supermarket models with repairable
servers, we provide a simple outline of the proof of Theorem \ref{The:Limit}.
To that end, it is a key to use the operator semigroup to provide a mean-field
limit for the sequence of Markov processes who asymptotically approaches a
single trajectory identified by the unique and global solution to an
infinite-dimensional system of mean-field equations. Readers may refer to Li
et al. \cite{Li:2014a} for more details with respect to the proof of such a
mean-field limit.

For the vector $\mathbf{u}^{\left(  N\right)  }=\left(  \mathbf{u}%
_{W}^{\left(  N\right)  },\mathbf{u}_{R}^{\left(  N\right)  }\right)  $ where
$\mathbf{u}_{W}^{\left(  N\right)  }=\left(  u_{0}^{\left(  N\right)  }%
,u_{1}^{\left(  N\right)  },u_{2}^{\left(  N\right)  },\ldots\right)  $ and
$\mathbf{u}_{R}^{\left(  N\right)  }=\left(  v_{1}^{\left(  N\right)  }%
,v_{2}^{\left(  N\right)  },v_{3}^{\left(  N\right)  },\ldots\right)  $, we
write%
\begin{align*}
\widetilde{\Omega}_{N}=  &  \left\{  \mathbf{u}^{\left(  N\right)  }=\left(
\mathbf{u}_{W}^{\left(  N\right)  },\mathbf{u}_{R}^{\left(  N\right)
}\right)  :1\geq u_{0}^{\left(  N\right)  }\geq u_{1}^{\left(  N\right)  }\geq
u_{2}^{\left(  N\right)  }\geq u_{3}^{\left(  N\right)  }\geq\cdots
\geq0,\right. \\
&  1\geq v_{1}^{\left(  N\right)  }\geq v_{2}^{\left(  N\right)  }\geq
v_{3}^{\left(  N\right)  }\geq v_{4}^{\left(  N\right)  }\geq\cdots\geq0,\\
&  \left.  Nu_{k}^{\left(  N\right)  }\text{ and }Nv_{l}^{\left(  N\right)
}\text{ are nonnegative integers for }k\geq0\text{ and }l\geq1\right\}
\end{align*}
and%
\[
\Omega_{N}=\left\{  \mathbf{u}^{\left(  N\right)  }\in\widetilde{\Omega}%
_{N}:\mathbf{u}^{\left(  N\right)  }e<+\infty\right\}  .
\]
At the same time, for the vector $\mathbf{u}=\left(  \mathbf{u}_{W}%
,\mathbf{u}_{R}\right)  $ where $\mathbf{u}_{W}=\left(  u_{0},u_{1}%
,u_{2},\ldots\right)  $ and $\mathbf{u}_{R}=\left(  v_{1},v_{2},v_{3}%
,\ldots\right)  $, we set%
\[
\widetilde{\Omega}=\{\mathbf{u}=\left(  \mathbf{u}_{W},\mathbf{u}_{R}\right)
:1\geq u_{0}\geq u_{1}\geq\cdots\geq0,1\geq v_{1}\geq v_{2}\geq\cdots\}
\]
and%
\[
\Omega=\left\{  \mathbf{u}\in\widetilde{\Omega}:\mathbf{u}e<+\infty\right\}
.
\]
Obviously, $\Omega_{N}\subsetneqq\Omega\subsetneqq\widetilde{\Omega}$ and
$\Omega_{N}\subsetneqq\widetilde{\Omega}_{N}\subsetneqq\widetilde{\Omega}$.

In the infinite-dimensional vector space $\widetilde{\Omega}$, we take a
metric%
\begin{equation}
\rho\left(  \mathbf{u},\mathbf{u}^{\prime}\right)  =\sup_{k\geq0,l\geq
1}\left\{  \dfrac{|u_{k}-u_{k}^{\prime}|}{k+1},\dfrac{|v_{l}-v_{l}^{\prime}%
|}{l}\right\}  , \label{Equat8}%
\end{equation}
for $\mathbf{u},\mathbf{u}^{\prime}\in\widetilde{\Omega}$. Note that under the
metric $\rho\left(  \mathbf{u},\mathbf{u}^{\prime}\right)  ,$ the
infinite-dimensional vector space $\widetilde{\Omega}$ is complete, separable
and compact.

For simplicity of description, here we only study the sequence $\left\{
\mathbf{U}^{\left(  N\right)  }\left(  t\right)  ,t\geq0\right\}  $ of Markov
processes in the first supermarket model with repairable servers, while the
other three models can be analyzed similarly without any difficulty.

For the first supermarket model with repairable servers, the Markov process
$\left\{  \mathbf{U}^{\left(  N\right)  }\left(  t\right)  ,t\geq0\right\}  $
is described as%
\[
\frac{\text{d}}{\text{d}t}\mathbf{U}^{\left(  N\right)  }\left(  t\right)
=\mathbf{A}_{N}\text{ }f(\mathbf{U}^{\left(  N\right)  }\left(  t\right)  ),
\]
where $\mathbf{A}_{N}$ acting on functions $f:\Omega_{N}\rightarrow
\mathbf{C}^{1}$ is the generating operator of the Markov process $\left\{
\mathbf{U}^{\left(  N\right)  }\left(  t\right)  ,t\geq0\right\}  $, and%
\begin{equation}
\mathbf{A}_{N}=\mathbf{A}_{N}^{\text{Input}}+\mathbf{A}_{N}^{\text{Out}%
}+\mathbf{A}_{N}^{\text{Transition}}, \label{Equat9}%
\end{equation}
for $\mathbf{u=}\left(  \mathbf{g,h}\right)  $, $\mathbf{g}=\left(
g_{0},g_{1},g_{2},\ldots\right)  $ and $\mathbf{h}=\left(  h_{1},h_{2}%
,h_{3},\ldots\right)  ,$%
\begin{align*}
\mathbf{A}_{N}^{\text{Input}}=  &  \lambda N\left(  g_{0}-g_{1}\right)
L_{1}\left(  g_{0},g_{1};h_{1}\right)  \left[  f(\mathbf{g}+\dfrac
{\mathbf{e}_{1}}{N},\mathbf{h})-f(\mathbf{g},\mathbf{h})\right] \\
&  +\lambda N\sum\limits_{k=2}^{\infty}\left\{  \left(  g_{k-1}-g_{k}\right)
L_{k}\left(  g_{k-1},g_{k};h_{k-1},h_{k}\right)  \left[  f(\mathbf{g}%
+\dfrac{\mathbf{e}_{k}}{N},\mathbf{h})-f(\mathbf{g},\mathbf{h})\right]
\right. \\
&  +\left.  \left(  h_{k-1}-h_{k}\right)  L_{k}\left(  g_{k-1},g_{k}%
;h_{k-1},h_{k}\right)  \left[  f(\mathbf{g},\mathbf{h}+\dfrac{\mathbf{e}_{k}%
}{N})-f(\mathbf{g},\mathbf{h})\right]  \right\}  ,
\end{align*}%
\[
\mathbf{A}_{N}^{\text{Out}}=\mu N\sum_{k=1}^{\infty}\left(  g_{k}%
-g_{k+1}\right)  \left[  f\left(  \mathbf{g}-\frac{\mathbf{e}_{k\text{ }}}%
{N},\mathbf{h}\right)  -f\left(  \mathbf{g,h}\right)  \right]
\]
and%
\begin{align*}
\mathbf{A}_{N}^{\text{Transition}}=  &  \alpha N\sum_{k=1}^{\infty}%
g_{k}\left[  f\left(  \mathbf{g}-\frac{\mathbf{e}_{k\text{ }}}{N}%
,\mathbf{h}+\frac{\mathbf{e}_{k\text{ }}}{N}\right)  -f\left(  \mathbf{g,h}%
\right)  \right] \\
&  +\beta N\sum_{k=1}^{\infty}h_{k}\left[  f\left(  \mathbf{g}%
+\frac{\mathbf{e}_{k\text{ }}}{N},\mathbf{h}-\frac{\mathbf{e}_{k\text{ }}}%
{N}\right)  -f\left(  \mathbf{g,h}\right)  \right]  ,
\end{align*}
where $\mathbf{e}_{k\text{ }}$stands for a row vector with the $k$th entry be
one and all the other entries be zero, and%
\[
L_{1}\left(  g_{0};g_{1},h_{1}\right)  =\sum_{m=1}^{d_{1}}C_{d_{1}}^{m}\left(
g_{0}-g_{1}\right)  ^{m-1}\left(  g_{1}+h_{1}\right)  ^{d-m},
\]
for $k\geq2$%
\[
L_{k}\left(  g_{k-1},g_{k};h_{k-1},h_{k}\right)  =\sum_{m=1}^{d_{1}}C_{d_{1}%
}^{m}\left(  g_{k-1}-g_{k}+h_{k-1}-h_{k}\right)  ^{m-1}\left(  g_{k}%
+h_{k}\right)  ^{d-m}.
\]
Therefore, for $\mathbf{u=}(\mathbf{g},\mathbf{h})\in\Omega_{N}$ and the
function $f:\Omega_{N}\rightarrow\mathbf{C}^{1}$ we obtain%
\begin{align}
\mathbf{A}_{N}f\left(  \mathbf{g,h}\right)  =  &  \lambda N\left(  g_{0}%
-g_{1}\right)  L_{1}\left(  g_{0},g_{1};h_{1}\right)  \left[  f(\mathbf{g}%
+\dfrac{\mathbf{e}_{1}}{N},\mathbf{h})-f(\mathbf{g},\mathbf{h})\right]
\nonumber\\
&  +\lambda N\sum\limits_{k=2}^{\infty}\left\{  \left(  g_{k-1}-g_{k}\right)
L_{k}\left(  g_{k-1},g_{k};h_{k-1},h_{k}\right)  \left[  f(\mathbf{g}%
+\dfrac{\mathbf{e}_{k}}{N},\mathbf{h})-f(\mathbf{g},\mathbf{h})\right]
\right. \nonumber\\
&  +\left.  \left(  h_{k-1}-h_{k}\right)  L_{k}\left(  g_{k-1},g_{k}%
;h_{k-1},h_{k}\right)  \left[  f(\mathbf{g},\mathbf{h}+\dfrac{\mathbf{e}_{k}%
}{N})-f(\mathbf{g},\mathbf{h})\right]  \right\} \nonumber\\
&  +\mu N\sum_{k=1}^{\infty}\left(  g_{k}-g_{k+1}\right)  \left[  f\left(
\mathbf{g}-\frac{\mathbf{e}_{k\text{ }}}{N},\mathbf{h}\right)  -f\left(
\mathbf{g,h}\right)  \right] \nonumber\\
&  +\alpha N\sum_{k=1}^{\infty}g_{k}\left[  f\left(  \mathbf{g}%
-\frac{\mathbf{e}_{k\text{ }}}{N},\mathbf{h}+\frac{\mathbf{e}_{k\text{ }}}%
{N}\right)  -f\left(  \mathbf{g,h}\right)  \right] \nonumber\\
&  +\beta N\sum_{k=1}^{\infty}h_{k}\left[  f\left(  \mathbf{g}%
+\frac{\mathbf{e}_{k\text{ }}}{N},\mathbf{h}-\frac{\mathbf{e}_{k\text{ }}}%
{N}\right)  -f\left(  \mathbf{g,h}\right)  \right]  . \label{Equat10}%
\end{align}

The operator semigroup of the Markov process $\left\{  \mathbf{U}%
^{(N)}(t),t\geq0\right\}  $ is defined as $\mathbf{T}_{N}(t)$, where if
$f:\Omega_{N}\rightarrow\mathbf{C}^{1}$, then for $\left(  \mathbf{g,h}%
\right)  \in\Omega_{N}$ and $t\geq0$%
\[
\mathbf{T}_{N}(t)f(\mathbf{g,h)}=E\left[  f(\mathbf{U}_{N}(t)\text{ }|\text{
}\mathbf{U}_{N}(0)=\left(  \mathbf{g,h}\right)  \right]  .
\]
Note that $\mathbf{A}_{N}$ is the generating operator of the operator
semigroup $\mathbf{T}_{N}(t)$, it is easy to see that $\mathbf{T}_{N}%
(t)=\exp\left\{  \mathbf{A}_{N}t\right\}  $ for $t\geq0$.

To analyze the limiting behavior of the sequence $\{\mathbf{U}^{\left(
N\right)  }\left(  t\right)  $, $t\geq0\}$ of the Markov processes, two formal
limits for the sequence $\left\{  \mathbf{A}_{N}\right\}  $ of the generating
operators and for the sequence $\left\{  \mathbf{T}_{N}(t)\right\}  $ of the
semigroups are expressed as $\mathbf{A}=\lim_{N\rightarrow\infty}%
\mathbf{A}_{N}$ and $\mathbf{T}\left(  t\right)  =\lim_{N\rightarrow\infty
}\mathbf{T}_{N}(t)$ for $t\geq0$, respectively. It follows from (\ref{Equat10}%
) that as $N\rightarrow\infty$%
\begin{align}
\mathbf{A}f\left(  \mathbf{g,h}\right)  =  &  \lambda N\left(  g_{0}%
-g_{1}\right)  L_{1}\left(  g_{0},g_{1};h_{1}\right)  \frac{\partial}{\partial
g_{1}}f(\mathbf{g},\mathbf{h})\nonumber\\
&  +\lambda N\sum\limits_{k=2}^{\infty}\left[  \left(  g_{k-1}-g_{k}\right)
L_{k}\left(  g_{k-1},g_{k};h_{k-1},h_{k}\right)  \frac{\partial}{\partial
g_{k}}f(\mathbf{g},\mathbf{h})\right. \nonumber\\
&  +\left.  \left(  h_{k-1}-h_{k}\right)  L_{k}\left(  g_{k-1},g_{k}%
;h_{k-1},h_{k}\right)  \frac{\partial}{\partial h_{k}}f(\mathbf{g}%
,\mathbf{h})\right] \nonumber\\
&  -\mu N\sum_{k=1}^{\infty}\left(  g_{k}-g_{k+1}\right)  \frac{\partial
}{\partial g_{k}}f(\mathbf{g},\mathbf{h})\nonumber\\
&  -\alpha N\sum_{k=1}^{\infty}g_{k}\left[  \dfrac{\partial}{\partial g_{k}%
}f(\mathbf{g},\mathbf{h})-\dfrac{\partial}{\partial h_{k}}f(\mathbf{g}%
,\mathbf{h})\right] \nonumber\\
&  +\beta N\sum_{k=1}^{\infty}h_{k}\left[  \dfrac{\partial}{\partial g_{k}%
}f(\mathbf{g},\mathbf{h})-\dfrac{\partial}{\partial h_{k}}f(\mathbf{g}%
,\mathbf{h})\right]  . \label{Equat12}%
\end{align}

The following theorem applies the operator semigroup to provide the mean-field
limiting process $\left\{  \mathbf{U}\left(  t\right)  ,t\geq0\right\}  $ for
the sequence $\left\{  \mathbf{U}^{\left(  N\right)  }\left(  t\right)
,t\geq0\right\}  $ of Markov processes, and indicates that this sequence of
Markov processes asymptotically approaches a single trajectory identified by
the unique and global solution to the system of mean-field equations. This
proof is omitted here. Readers may refer to Li et al. \cite{Li:2014a} for more details.

\begin{The}
\label{The:Lim}Let $f$ be continuous functions $f:\widetilde{\Omega
}\rightarrow\mathbf{C}^{1}$. Then for any $t>0$%
\[
\lim_{N\rightarrow\infty}\underset{(\mathbf{g},\mathbf{h)}\in\Omega_{N}}{\sup
}\left\vert \mathbf{T}_{N}\left(  t\right)  f\left(  \mathbf{g},\mathbf{h}%
\right)  -f\left(  \mathbf{u}(t,\mathbf{g},\mathbf{h})\right)  \right\vert
=0.
\]
The convergence is uniform in $t\in\left[  0,a\right]  $ for any $a>0.$
\end{The}

Finally, we provide some interpretation on Theorem \ref{The:Lim}. If
$\lim_{N\rightarrow\infty}\mathbf{U}^{\left(  N\right)  }\left(  0\right)
=\mathbf{u}(0)=\left(  \mathbf{g},\mathbf{h}\right)  \in$ $\Omega$ in
probability, then Theorem \ref{The:Lim} shows that $\mathbf{U}\left(
t\right)  =\lim_{N\rightarrow\infty}\mathbf{U}^{\left(  N\right)  }\left(
t\right)  $ is concentrated on the trajectory $\Gamma_{\mathbf{g}}=\left\{
\mathbf{u}(t,\mathbf{g,h}):t\geq0\right\}  $. This indicates the functional
strong law of large numbers for the time evolution of the fraction of each
state of this supermarket model, thus the sequence $\left\{  \mathbf{U}%
^{\left(  N\right)  }\left(  t\right)  ,t\geq0\right\}  $ of Markov processes
converges weakly to the expected fraction vector $\mathbf{u}(t,\mathbf{g}%
,\mathbf{h})$ as $N\rightarrow\infty$, that is, for any $T>0$%
\[
\lim_{N\rightarrow\infty}\sup_{0\leq s\leq T}\left\|  \mathbf{U}^{\left(
N\right)  }\left(  s\right)  -\mathbf{u}(s,\mathbf{g},\mathbf{h})\right\|
=0\text{ \ in probability}.
\]
Note that the limits are necessary for using the stationary probabilities of
the limiting process to give an effective approximate performance of this
supermarket model.

\textbf{The Proof of Theorem \ref{The:Limit}}

In the remainder of this Appendix, we discuss some useful limits of the
fraction vector $\mathbf{u}^{\left(  N\right)  }\left(  t\right)  $ as
$N\rightarrow\infty$ and $t\rightarrow+\infty$ whose purpose is to give the
proof of Theorem \ref{The:Limit}.

The following theorem gives the limit of the vector $\mathbf{u}(t,\mathbf{g}%
,\mathbf{h})$ as $t\rightarrow+\infty$, that is,%
\[
\lim_{t\rightarrow+\infty}\mathbf{u}(t,\mathbf{g},\mathbf{h})=\lim
_{t\rightarrow+\infty}\lim_{N\rightarrow\infty}\mathbf{u}^{\left(  N\right)
}(t,\mathbf{g},\mathbf{h}).
\]
This proof is omitted here. Readers may refer to Li et al. \cite{Li:2014a} for
more details.

\begin{The}
\label{The:Limit1}If $\widetilde{\rho}<1$, then for any $\left(
\mathbf{g},\mathbf{h}\right)  \in\Omega$%
\[
\lim_{t\rightarrow+\infty}\mathbf{u}(t,\mathbf{g},\mathbf{h})=\pi.
\]
Furthermore, there exists a unique probability measure $\varphi$ on $\Omega$,
which is invariant under the map $\left(  \mathbf{g},\mathbf{h}\right)
\longmapsto\mathbf{u}(t,\mathbf{g},\mathbf{h})$, that is, for any continuous
function $f:$ $\Omega\rightarrow\mathbf{R}$ and $t>0$%
\[
\int_{\Omega}f(\mathbf{g},\mathbf{h})\text{d}\varphi(\mathbf{g},\mathbf{h}%
)=\int_{\Omega}f(\mathbf{u}(t,\mathbf{g},\mathbf{h}))\text{d}\varphi
(\mathbf{g},\mathbf{h}).
\]
Also, $\varphi=\delta_{\pi}$ is the probability measure concentrated at the
fixed point $\pi$.
\end{The}

The following theorem indicates the weak convergence of the sequence $\left\{
\varphi_{N}\right\}  $ of stationary probability distributions for the
sequence $\left\{  \mathbf{U}^{\left(  N\right)  }(t),t\geq0\right\}  $ of
Markov processes to the probability measure concentrated at the fixed point
$\pi$. This proof is omitted here. Readers may refer to Li et al.
\cite{Li:2014a} for more details.

\begin{The}
\label{The:Limit2}(1) If $\widetilde{\rho}<1$, then for a fixed number
$N=1,2,3,\ldots$, the Markov process $\left\{  \mathbf{U}^{\left(  N\right)
}(t),t\geq0\right\}  $ is positive recurrent, and has a unique invariant
distribution $\varphi_{N}$.

(2) $\left\{  \varphi_{N}\right\}  $ weakly converges to $\delta_{\pi}$, that
is, for any continuous function $f:$ $\Omega\rightarrow\mathbf{R}$%
\[
\lim_{N\rightarrow\infty}E_{\varphi_{N}}\left[  f(\mathbf{g},\mathbf{h}%
)\right]  =f\left(  \pi\right)  .
\]
\end{The}

Based on Theorems \ref{The:Limit1} and \ref{The:Limit2}, we obtain a useful
relation as follows%
\[
\lim_{t\rightarrow+\infty}\lim_{N\rightarrow\infty}\mathbf{u}^{\left(
N\right)  }(t,\mathbf{g},\mathbf{h})=\lim_{N\rightarrow\infty}\lim
_{t\rightarrow+\infty}\mathbf{u}^{\left(  N\right)  }(t,\mathbf{g}%
,\mathbf{h})=\pi.
\]
Therefore, we have%
\[
\lim_{\substack{N\rightarrow\infty\\t\rightarrow+\infty}}\mathbf{u}^{\left(
N\right)  }(t,\mathbf{g},\mathbf{h})=\pi.
\]
Clearly, the above analysis completes the proof of Theorem \ref{The:Limit}.

\vskip                                                                  0.2cm

\end{document}